\documentclass[aps,prx,twocolumn,superscriptaddress,showpacs,longbibliography]{revtex4-1}

\usepackage{amsmath}
\usepackage{amsfonts}
\usepackage{amssymb}
\usepackage{graphicx}
\usepackage{color}
\usepackage{bm}
\usepackage{hyperref}
\usepackage[english]{babel}

\newcommand{\av}[1]{\left<#1\right>}

\newcommand{\sgn}{\mathop{\mathrm{sgn}}}
\newcommand{\diag}{\mathop{\mathrm{diag}}}

\begin{document}

\title{Bulk Pumping in 2D Topological Phases}

\author{Charles-Edouard Bardyn}
\affiliation{Department of Quantum Matter Physics, University of Geneva, 24 Quai Ernest-Ansermet, CH-1211 Geneva, Switzerland}
\author{Michele Filippone}
\affiliation{Department of Quantum Matter Physics, University of Geneva, 24 Quai Ernest-Ansermet, CH-1211 Geneva, Switzerland}
\author{Thierry Giamarchi}
\affiliation{Department of Quantum Matter Physics, University of Geneva, 24 Quai Ernest-Ansermet, CH-1211 Geneva, Switzerland}


\begin{abstract}
The notion of topological (Thouless) pumping in topological phases is traditionally associated with Laughlin's pump argument for the quantization of the Hall conductance in two-dimensional (2D) quantum Hall systems. It relies on magnetic flux variations that thread the system of interest without penetrating its bulk, in the spirit of Aharonov-Bohm effects. Here we explore a different paradigm for topological pumping induced, instead, by magnetic flux variations $\delta\chi$ inserted through the bulk of topological phases. We show that $\delta\chi$ generically controls the analog of a topological pump, accompanied by robust physical phenomena. We demonstrate this concept of bulk pumping in two paradigmatic types of 2D topological phases: integer and fractional quantum Hall systems and topological superconductors. We show, in particular, that bulk pumping provides a unifying connection between seemingly distinct physical effects such as density variations described by Streda's formula in quantum Hall phases, and fractional Josephson currents in topological superconductors. We discuss the generalization of bulk pumping to other types of topological phases.
\end{abstract}


\maketitle



The notion of topological pump introduced by Thouless~\cite{Thouless1983} underlies some of the most robust quantum phenomena. In essence, it corresponds to the dynamical implementation of a gauge transformation, via deformations of the Hamiltonian of a quantum system in some parameter space. In a topological pump, the net result of a parameter cycle is nontrivial: Though the Hamiltonian remains identical up to the applied gauge transformation, a permutation occurs between the system's eigenstates, leading to robust (typically quantized) physical effects.

\begin{figure}[t]
\begin{center}
    \includegraphics[width=\columnwidth]{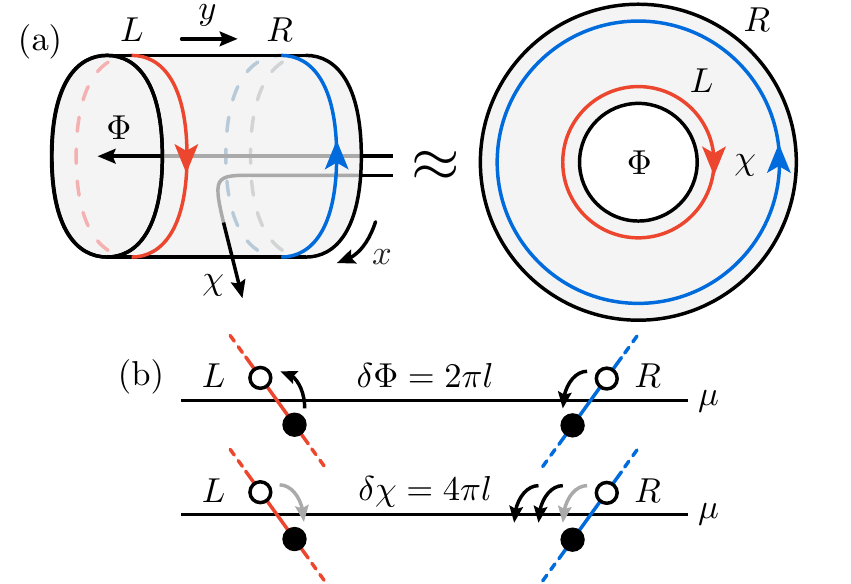}
    \caption{\textbf{(a)}: Schematic setup for conventional (Laughlin's) topological pumping vs. ``bulk pumping''. A gapped topological phase with cylinder or Corbino disk geometry is exposed to two types of magnetic fluxes: an Aharonov-Bohm flux $\Phi$ threading its hole, and a transverse flux $\chi$ threading its bulk. In the examples considered here, phases exhibit counterpropagating gapless edge modes [red (L) and blue (R)] supporting quasiparticle excitations with charge $e/l$, where $l \geq 1$ is an odd integer. \textbf{(b)}: Schematic anomalous spectral flow of low-energy (edge) modes due to small flux variations $\delta\Phi$ and $\delta\chi$. While Laughlin's conventional pump corresponds to $\delta\Phi = 2\pi l$ ($l$ flux quanta)~\cite{Laughlin1981,Halperin1982}, the bulk pump of interest corresponds to $\delta\chi = 4\pi l$ (see text). Bulk pumping induces spectral flow at the right edge only [or opposite spectral flows at opposite edges (shown in grey), in a gauge where $\delta\Phi \to \delta\Phi + \delta\chi/2$; see Eq.~\eqref{eq:gaugeChoice2}]. Empty/filled dots represent empty/filled states (see text), and $\mu$ denotes the Fermi level energy.}
    \label{fig:setup}
\end{center}
\end{figure}

Magnetic fluxes offer a natural ``knob'' to induce interesting pumping effects. One of the most famous examples is provided by Laughlin's argument~\cite{Laughlin1981,Halperin1982}, which relates the variation of a magnetic flux to a quantized charge-pumping effect: the Hall conductance of two-dimensional (2D) quantum Hall systems. Laughlin's pump paradigm corresponds to the situation where all system's eigenstates undergo a flux-induced circular shift in momentum space (e.g., around a crystal Brillouin zone).

While Laughlin's pump relies on variations $\delta\Phi$ of an Aharonov-Bohm flux $\Phi$ (threading the system without penetrating its bulk, as in Fig.~\ref{fig:setup}a), variations $\delta\chi$ of a magnetic flux $\chi$ inserted \emph{through the bulk} of topological phases can also give rise to robust phenomena: In 2D quantum Hall phases, e.g., transverse flux variations $\delta\chi$ induce density changes proportional to the quantized Hall conductance, as described by Streda's formula~\cite{Streda1982}. In 2D topological superconductors, in contrast, the insertion of flux quanta through the bulk gives rise to the creation or annihilation (fusion) of Majorana zero modes~\cite{Read2000,Kitaev2001,Kitaev2006}, and to fractional Josephson effects~\cite{Kitaev2001,Fu2008,Fu2009,Lindner2012,Cheng2012,Clarke2013} (see, e.g., Refs.~\cite{Nayak2008,Alicea2012} for reviews). Despite their topological origins, such effects are conventionally derived and understood on a case-by-case basis, without apparent connection to topological pumping.

In this work, we demonstrate that the insertion of flux quanta $\delta\chi$ through the bulk of gapped topological phases with protected gapless edge states generically leads to robust pumping effects. We relate $\delta\chi$ to the low-energy analog of a topological pump, which we coin ``bulk pump'', and argue that robust phenomena induced by $\delta\chi$ can be understood and systematically be searched for using this notion of bulk pumping. We demonstrate our claims in two paradigmatic types of topological phases: 2D quantum Hall systems and 2D topological superconductors. We show that effects associated with Streda's formula and fractional Josephson currents are manifestations of the same bulk pump $\delta\chi$ in distinct topological phases, and argue that bulk pumping provides a generic knob for robust effects in topological phases.

\section{Key results and outline}

Our goal is to demonstrate that the insertion of flux quanta through the bulk of gapped phases with topologically protected gapless edge states induces the analog of a topological pump, accompanied by robust physical effects. To highlight the similarities and differences between bulk and conventional topological pumping, we consider a similar setup as in Laughlin's argument: 2D gapped topological phases on a cylinder (or, equivalently for our purposes, on a Corbino disk), as shown in Fig.~\ref{fig:setup}a. We consider phases whose low-energy properties are governed by robust gapless edge modes appearing at the edges of the cylinder, and focus on phases made of fermions with charge $e = 1$, minimally coupled to two types of external magnetic fluxes: (i) an Aharonov-Bohm flux $\Phi$ threading the system's hole --- controlling Laughlin's pump --- and (ii) a transverse flux $\chi$ inserted through the bulk of the system --- controlling the bulk pump of interest. We assume that fermions are spinless (or spin-polarized), and set the system's temperature to zero. Our results are readily extendable to finite temperature and more complicated types of geometries.

Our paper is organized as follows: In Sec.~\ref{sec:bulkPumpQH}, we demonstrate the effects of bulk pumping in paradigmatic examples of noninteracting (integer) and interacting (fractional) 2D quantum Hall phases~\cite{Laughlin1983,Halperin1983,Jain1989}.
Focusing on low-energy (edge) modes, we show that flux variations $\delta\Phi$ and $\delta\chi$ control distinct types of chiral anomalies~\cite{Adler1969,Bell1969}: While $\delta\Phi$ (Laughlin's pump) induces a global momentum shift or unidirectional spectral flow of all system's eigenstates, $\delta\chi$ essentially induces opposite spectral flows at opposite edges (see Fig.~\ref{fig:setup}b). The physical effects of $\delta\Phi$ and $\delta\chi$ are thus very distinct yet similarly robust. In the quantum Hall phases of interest, we show that $\delta\chi$ controls the pumping of charges from the edges into the bulk, or vice versa, in agreement with Streda's formula~\cite{Streda1982}. In particular, the insertion of $l$ bulk flux quanta, $\delta\chi = 2\pi l$ (in natural units $\hbar = c = e = 1$, where $e/l$ is the charge of underlying quasiparticle excitations), leads to an apparent fermion-number parity switch.
A similar effect was recently identified for persistent currents in noninteracting mesoscopic quantum ladders (where $l = 1$)~\cite{Filippone2017}.

In Sec.~\ref{sec:bulkPumpSC}, we extend our discussion to superconducting analogs of the integer and fractional quantum Hall phases considered in Sec.~\ref{sec:bulkPumpQH}~\cite{Read2000,Kitaev2001,Lindner2012,Cheng2012,Clarke2013}. We construct an effective field-theory description of low-energy edge modes in the same vein as conventional edge theories for quantum Hall phases. We then demonstrate that
parity conservation in superconductors leads to pumping effects that exhibit a robust $4\pi l$ periodicity in $\delta\chi$. We explicitly relate the insertion of bulk flux quanta to fractional Josephson effects~\cite{Kitaev2001,Fu2008,Fu2009,Lindner2012,Cheng2012,Clarke2013}.

We present our conclusions in Sec.~\ref{sec:conclusions}, and provide additional information and theoretical background in three Appendices where the 2D topological phases examined in the main text are described using coupled 1D wires, in the spirit of Refs.~\cite{Kane2002,Teo2014}: In Appendices~\ref{sec:tightBindingQH} and~\ref{sec:tightBindingSC}, we detail the effects of bulk pumping in explicit tight-binding models for 2D integer ($l = 1$) quantum Hall and topological superconducting phases. In Appendix~\ref{sec:generalizedCoupledWireConstruction}, we present explicit derivations of the effective field theories used to describe edge modes in the main text. Our construction follows along the lines of Refs.~\cite{Neupert2014,Huang2016}, based on a formulation of Abelian bosonization by Haldane~\cite{Haldane1995}.

\section{Quasi-topological bulk pump in quantum Hall phases}
\label{sec:bulkPumpQH}

We start by examining the paradigmatic example of Abelian quantum Hall phases with filling factor $\nu = 1/l$, where $l \geq 1$ is an odd integer~\cite{Laughlin1983,Halperin1983,Jain1989} [details of the construction and properties of such phases can be found in Appendices~\ref{sec:tightBindingQH} (tight-binding coupled-wire picture for $l = 1$) and~\ref{sec:bosonizedPictureQH} (generalized bosonized picture for $l \geq 1$)]. The system has a global $U(1)$ symmetry reflecting charge conservation. Its low-energy physics is governed by a pair of counterpropagating chiral gapless edge modes, as in Fig~\ref{fig:setup}a. A uniform transverse field $\chi = \chi_0$ (whose value is irrelevant here) is required to generate the phase, and we set $\Phi = 0$, without loss of generality. In the absence of additional flux variations $\delta\Phi$ and $\delta\chi$, gapless edge modes are described by the Hamiltonian
\begin{equation} \label{eq:edgeHamiltonianQH}
    H_\sigma = \frac{v_\sigma l}{4\pi} \int dx (\partial_x \varphi_\sigma)^2,
\end{equation}
where $\sigma = -/+ \equiv L/R$ identifies the left/right edge of the system and the corresponding left/right chirality of the edge modes with velocity $v_\sigma$ (see Fig~\ref{fig:setup}a). The fields $\varphi_\sigma \equiv \varphi_\sigma(t, x)$ are chiral bosonic fields. 
Their chiral nature comes from their equal-time commutation relations
\begin{align}
    \left[ \varphi_\sigma(x), \varphi_\sigma(x') \right] & = -\sigma (i\pi/l) \sgn(x-x'), \label{eq:KacMoodyQH1} \\
    \left[ \varphi_L(x), \varphi_R(x') \right] & = (i\pi/l^2), \label{eq:KacMoodyQH2}
\end{align}
forming a $U(1)$ Kac-Moody algebra at level $l$. The second commutator arises from Klein factors, with conventions detailed in Secs.~\ref{sec:mainFramework} and~\ref{sec:bosonizedPictureQH} of Appendix~\ref{sec:generalizedCoupledWireConstruction}~\footnote{The edge fields $\varphi_\sigma(x)$ satisfy similar algebraic properties as chiral fields in a conventional Luttinger liquid. This can be seen by identifying the integer $l$ with the inverse Luttinger parameter, i.e., $l = 1/K$. Note that our conventions for the commutator in Eq.~\eqref{eq:KacMoodyQH1} differ from that of Ref.~\cite{Giamarchi2003} [Eq.~(3.59) thereof] by a sign. This corresponds to a change of coordinates $x \to -x$.}. The fields $\varphi_\sigma$ satisfy periodic boundary conditions
\begin{equation} \label{eq:periodicBCPhi}
    \varphi_\sigma(x + L_x) = \varphi_\sigma(x) + 2\pi n_\sigma,
\end{equation}
where $n_\sigma$ is an integer, 
and $L_x$ is the length of the system in the $x$ (azimuthal) direction.

Equations~\eqref{eq:edgeHamiltonianQH}-\eqref{eq:periodicBCPhi} describe quasiparticles propagating along each of the edges $\sigma$ with velocity $v_\sigma l$ and chirality $\sigma$. These so-called ``Laughlin quasiparticles'' are created by operators proportional to the normal-ordered vertex operators $(\Psi_\sigma^\text{qp})^\dagger = \exp[-i \varphi_\sigma]$. 
They carry a charge $e/l$, and exhibit a phase $e^{i\pi/l}$ under spatial exchange (see Appendix~\ref{sec:bosonizedPictureQH}). Operators $(\Psi_\sigma^\text{f})^\dagger = \exp[-i l \varphi_\sigma] = [(\Psi_\sigma^\text{qp})^\dagger]^l$ create fermions with unit charge, corresponding to ``ensembles'' of $l$ Laughlin quasiparticles.

When introducing flux variations $\delta\Phi$ and $\delta\chi$, the low-energy edge theory $H_L + H_R$ described by Eqs.~\eqref{eq:edgeHamiltonianQH}-\eqref{eq:periodicBCPhi} is modified in two ways: First, the edge-mode velocities $v_\sigma$ change by a nonuniversal value of order $\delta\chi/S$, where $S$ is the surface area of the system (see Appendix~\ref{sec:tightBindingQH}). Here, however, we neglect such corrections by focusing on ``microscopic'' flux variations $\delta\chi \ll S$, corresponding to the insertion of a small number of bulk flux quanta over the whole system. Second, and most importantly, the minimal coupling between the system's charges and flux variations leads to the replacement
\begin{equation} \label{eq:minimalCouplingChiralBosonicFields}
    \varphi_\sigma(x) \to \varphi_\sigma(x) - \frac{e}{l} \int_0^x dx' \delta A_\sigma(x')
\end{equation}
in Eq.~\eqref{eq:edgeHamiltonianQH} (see Appendix~\ref{sec:bosonizedPictureQH}), where $\delta A_\sigma(x)$ is the value, at the edge $\sigma$, of the $U(1)$ gauge field describing both $\delta\Phi$ and $\delta\chi$. As can be seen in Fig~\ref{fig:setup}a, the left (inner) edge $\sigma = -$ experiences a flux $\Phi$, while the right (outer) edge $\sigma = +$ is threaded by a total flux $\Phi + \chi$. A natural choice of gauge is thus $\delta A_\sigma(x) \equiv \delta A_\sigma$ (uniform), with
\begin{align} \label{eq:gaugeChoice1}
\begin{split}
    \delta A_- & \equiv \delta A_L = \delta\Phi/L_x, \\
    \delta A_+ & \equiv \delta A_R = (\delta\Phi + \delta\chi)/L_x,
\end{split}
\end{align}
which is equivalent to the more symmetric expression
\begin{equation} \label{eq:gaugeChoice2}
    \delta A_\sigma = (\delta\Phi + \sigma \delta\chi/2)/L_x, \quad \delta\Phi \to \delta\Phi + \delta\chi/2.
\end{equation}
Note that $\delta\Phi$ can be described as a phase ``twist'' $e^{-i\delta\Phi}$ in the boundary conditions of the fermionic fields $\Psi_\sigma^\text{f}$~\cite{Niu1985}. In particular, the shift $\delta\Phi \to \delta\Phi + \delta\chi/2$ in Eq.~\eqref{eq:gaugeChoice2} is equivalent to a phase twist $e^{-i \delta\chi/2}$ for fermions, which corresponds, for chiral bosonic fields $\varphi_\sigma$, to modified (twisted) boundary conditions
\begin{equation} \label{eq:twistedBCPhi}
    \varphi_\sigma(x + L_x) = \varphi_\sigma(x) + \delta\chi/2 + 2\pi n_\sigma.
\end{equation}
The low-energy theory described by Eqs.~\eqref{eq:edgeHamiltonianQH}-\eqref{eq:periodicBCPhi} exhibits a chiral anomaly~\cite{Adler1969,Bell1969},
which plays a key role in this work: Under flux variations $\delta\Phi, \delta\chi$, the number of charges in individual edge modes (the number of fermions with fixed chirality) is not conserved. Specifically, the operator describing the total charge in mode $\varphi_\sigma$, given by $Q_\sigma = -\sigma e/(2\pi) \int dx \partial_x \varphi_\sigma$ (see Appendix~\ref{sec:bosonizedPictureQH}), only satisfies $\partial_t Q_\sigma = i[H_L + H_R, Q_\sigma] = 0$ when $\delta\Phi = \delta\chi = 0$. When flux variations are introduced, in contrast, $\partial_x \varphi_\sigma$ is replaced by its covariant analog $D_x \varphi_\sigma \equiv \partial_x \varphi_\sigma - (e/l) \delta A_\sigma$ [Eq.~\eqref{eq:minimalCouplingChiralBosonicFields}], and the conserved-charge operator becomes
\begin{equation} \label{eq:edgeConservedCharge}
    \tilde{Q}_\sigma = -\frac{\sigma e}{2\pi} \int dx D_x \varphi_\sigma = Q_\sigma + \frac{\sigma e}{2\pi l} \int dx \delta A_\sigma.
\end{equation}
Expressing $\partial_t \tilde{Q}_\sigma = 0$ in the gauge defined by Eq.~\eqref{eq:gaugeChoice2}, we thus find edge currents of the form
\begin{equation} \label{eq:anomalousEdgeCurrents}
    J_\sigma \equiv \partial_t Q_\sigma = -\frac{\sigma e}{2\pi l} \partial_t \left( \delta\Phi + \sigma \delta\chi/2 \right),
\end{equation}
with implicit shift $\delta\Phi \to \delta\Phi + \delta\chi/2$ as in Eq.~\eqref{eq:gaugeChoice2}. This expression captures the main behavior of the system under flux variations: It shows that both types of fluxes $\delta\Phi$ and $\delta\chi$ contribute to anomalous (nonzero) charge transfers $J_\sigma$ between the edge modes and the rest of the system (the bulk). Specifically, Aharonov-Bohm flux variations $\delta\Phi$ induce currents of opposite signs at opposite edges --- into the bulk at one edge, and out of the bulk at the other edge --- while bulk flux variations $\delta\chi$ induce currents of the same sign at both edges --- into or out of the bulk at both edges (see Fig.~\ref{fig:setup}b).

The above anomalies are also visible in the edge spectrum. Indeed, charge excitations composed of $1 \leq s \leq l$ edge Laughlin quasiparticles satisfy
\begin{equation} \label{eq:commutatorHPsiQp}
    \left[ H_\sigma, [(\Psi_\sigma^\text{qp})^\dagger]^s \right] = \sigma v_\sigma \left( -i \partial_x - q_s \delta A_\sigma \right) [(\Psi_\sigma^\text{qp})^\dagger]^s.
\end{equation}
Moving to momentum space by defining $[(\Psi_\sigma^\text{qp})^\dagger]^s(p) = \int dx e^{-ip x} [(\Psi_\sigma^\text{qp})^\dagger]^s(x)$, the corresponding energy dispersion reads
\begin{align}
    E_{\sigma, s}(p) &= \sigma v_\sigma \left( p - q_s \delta A_\sigma \right) \nonumber \\
    &= \sigma v_\sigma \left( p - q_s \frac{\delta\Phi + \sigma \delta\chi/2}{L_x} \right),\label{eq:anomalousEdgeSpectralFlow}
\end{align}
where $q_s = (s/l)e$ is the relevant charge, and $p$ is the (conserved) momentum in the $x$ direction ($p = 2\pi n/L_x$ with integer $n$). Equation~\eqref{eq:anomalousEdgeSpectralFlow} shows that flux variations $\delta\Phi, \delta\chi$ induce an anomalous spectral flow~\cite{Wen2007,Bernevig2013} consistent with the edge current $J_\sigma$ in Eq.~\eqref{eq:anomalousEdgeCurrents}: While $\delta\Phi$ generates energy shifts $E_{\sigma, s} \to E_{\sigma, s} - \sigma v_\sigma q_s \delta\Phi/L_x$ of opposite signs at opposite edges, $\delta\chi$ induces shifts $E_{\sigma, s} \to E_{\sigma, s} - v_\sigma q_s (\delta\chi/2)/L_x$ of the same sign at both edges [up to an additional global shift $E_{\sigma, s} \to E_{\sigma, s} - \sigma v_\sigma q_s (\delta\chi/2)/L_x$ due to $\delta\Phi \to \delta\Phi + \delta\chi/2$ in Eq.~\eqref{eq:gaugeChoice2}] (see Fig.~\ref{fig:setup}b).

According to Eq.~\eqref{eq:anomalousEdgeSpectralFlow}, $\delta\Phi = 2\pi l$ is the minimal flux variation that leaves the low-energy theory of the system invariant, acting as a gauge transformation on the latter. Formally, this corresponds to the only true [global U(1)] gauge symmetry of the system, responsible for charge conservation. In practice, however, $\delta\chi = 2\pi l$ also leaves the low-energy theory approximately invariant, up to negligible nonuniversal corrections of the edge-mode velocities $v_\sigma$ (see Appendix~\ref{sec:tightBindingQH}). 

The above discussion shows that flux variations $\delta\Phi$ and $\delta\chi$
couple to two types of anomalies: symmetric and antisymmetric currents
\begin{align}
    J_v \equiv \frac{1}{2} \left( J_R + J_L \right) &= -\frac{e}{4\pi l} \partial_t \delta\chi, \label{eq:vectorCurrent} \\
    J_a \equiv \frac{1}{2} \left( J_R - J_L \right) &= -\frac{e}{2\pi l} \partial_t \delta\Phi. \label{eq:axialCurrent}
\end{align}
We call these ``vector'' and ``axial'' currents, respectively, in accordance with seminal studies of chiral anomalies by Adler, Bell and Jackiw (ABJ) in the context of pion decay~\cite{Adler1969,Bell1969}, later extended to various condensed matter systems~\cite{Nielsen1983,Alekseev1998,Volovik2009,Zyuzin2012,Kharzeev2013,Chen2013,Basar2014,Landsteiner2014,Huang2016}. 

In our setup, the anomaly controlled by $\delta\Phi$ --- known as chiral, axial, or ABJ anomaly --- 
underpins Laughlin's charge-pumping argument~\cite{Laughlin1981,Halperin1982}: The axial current $J_a$ represents a charge transfer between the two edges of the system, corresponding to the standard Hall current. The insertion of $l$ flux quanta $\delta\Phi = 2\pi l$ is a topological pumping process whereby one charge is transferred between the edges [see Eqs.~\eqref{eq:anomalousEdgeCurrents},~\eqref{eq:anomalousEdgeSpectralFlow}, and Fig.~\ref{fig:setup}b]. This pump is ``topological'' for two reasons: (i) the corresponding anomaly outflows $J_L$ and $J_R$ are nonzero, which requires the bulk to be in a topological phase, and (ii) $J_L$ and $J_R$ exactly cancel out, implying that $\delta\Phi = 2\pi l$ is a topological pump in the sense of Thouless~\cite{Thouless1983}, i.e., a closed cycle in parameter space leaving the system invariant up a gauge transformation.

The situation is different for the anomaly controlled by the bulk flux $\delta\chi$ of interest here: The vector current $J_v$ induced by $\delta\chi$ describes a charge transfer from the edges into the bulk (or vice versa, when $\delta\chi < 0$). The insertion of $l$ bulk flux quanta, $\delta\chi = 2\pi l$, is a pumping process whereby exactly one charge is transferred from the edges into the bulk [see Eqs.~\eqref{eq:anomalousEdgeCurrents},~\eqref{eq:anomalousEdgeSpectralFlow}, and Fig.~\ref{fig:setup}b]. This bulk pump is topological in the sense that it relies on a topological bulk. In contrast to Laughlin's pump, however, it is not topological in the sense of Thouless: As mentioned above, $\delta\chi = 2\pi l$ does not represent a true gauge transformation of the full system. The anomaly outflows $J_L$ and $J_R$ do not cancel out (they add up), and the bulk must change in order to absorb the total outflow $J_L + J_R$, corresponding to an additional charge transferred from the edges. From the viewpoint of low-energy (edge) modes, however, bulk modifications only lead to corrections of order $\delta\chi/S \ll 1$ of the edge-mode velocities $v_\sigma$ (Appendix~\ref{sec:tightBindingQH}). Therefore, for low-energy phenomena, the only difference between $\delta\chi = 2\pi l$ and a true Thouless pump are small corrections $2\pi l/S$. Accordingly, we identify $\delta\chi = 2\pi l$ as a ``quasi-topological'' bulk pump.

We remark that the pumps $\delta\Phi = 2\pi l$ and $\delta\chi = 2\pi l$ transfer the same amount of charge despite their distinct physical and topological nature. This can be regarded as a manifestation of Streda's formula relating the Hall conductance induced by $\delta\Phi$ to bulk density changes induced by variations $\delta\chi$ of the transverse magnetic field~\cite{Streda1982}.

As we demonstrate in additional examples below, the insertion of bulk flux quanta $\delta\chi$ generically leads to robust pumping effects. The physical meaning and periodicity (in number of bulk flux quanta) of these effects depend on the nature of the underlying topological phase and, more importantly, on the corresponding anomalous low-energy edge theory. In the quantum Hall phases examined so far, $\delta\chi = 2\pi l$ induces an anomalous spectral flow where one of the two occupied edge fermionic modes at the Fermi level 
flows from the edges into the bulk, thereby pumping one charge into the latter. The number of bulk fermionic modes increases by one in the process. One can then distinguish two scenarios: (i) If the Fermi level is pinned by an external reservoir of charges, the total number of fermions in the system increases by one. (ii) If the total number of fermions, instead, is conserved, the pump leads to an apparent change of fermion-number parity: For $\delta\chi = 2\pi l$, one of the two occupied edge fermionic modes at the Fermi level is emptied, while for $\delta\chi = 4\pi l$ both are emptied, and the system comes back to a configuration with occupied edge fermionic modes and a lower Fermi level. This effective parity ``switch'' leads to an apparent $4\pi l$ periodicity (in $\delta\chi$) for phenomena that depend on parity. This could be observed, e.g., by measuring persistent currents in a mesoscopic system~\cite{Filippone2017}.

The fact that $\delta\chi = 4\pi l$ pumps exactly two fermionic modes (or charges) from the edges into the bulk, corresponding to a double bulk parity switch, hints at a way to obtain more robust pumping effects: If the $U(1)$ symmetry responsible for fermion-number conservation was broken down to a $\mathbb{Z}_2$ symmetry corresponding to fermion-number parity conservation, the bulk would be able to absorb pairs of fermions without breaking symmetries, which would promote $\delta\chi = 4\pi l$ to a bona fide low-energy topological pump. We demonstrate this below by extending our discussion to topological superconductors.

\section{Topological bulk pump in topological superconductors}
\label{sec:bulkPumpSC}

To examine the effects of bulk flux quanta in topological phases with $\mathbb{Z}_2$ fermion-number-parity conservation, we consider the closest superconducting analog of the quantum Hall phases examined so far: topological superconducting phases made of spinless fermions with unit charge, and protected by particle-hole (PH) symmetry alone (i.e., in symmetry class $D$ of conventional classifications~\cite{Zirnbauer1996,Altland1997} containing, e.g., 2D $p$-wave topological superconductors~\cite{Read2000}). Details regarding the construction and properties of such phases can be found in Appendices~\ref{sec:tightBindingSC} (tight-binding coupled-wire picture for $l = 1$) and~\ref{sec:bosonizedPictureSC} (generalized bosonized picture for $l \geq 1$). As no background transverse flux is required here, we start with $\Phi = \chi = 0$, without loss of generality. As detailed in Appendix~\ref{sec:bosonizedPictureSC}, the relevant low-energy physics is described by the following PH-symmetric analog of Eq.~\eqref{eq:edgeHamiltonianQH}:
\begin{equation} \label{eq:edgeHamiltonianSC}
    H_\sigma = \frac{v_\sigma l}{4\pi} \int dx \frac{1}{2} [(\partial_x \varphi_\sigma)^2 + (\partial_x \bar{\varphi}_\sigma)^2],
\end{equation}
where $\sigma$ and $\varphi_\sigma$ are defined as before, and $v_\sigma$ essentially corresponds, here, to the amplitude of superconducting pairings in the topological phase (see Appendix~\ref{sec:tightBindingSC}). Equation~\eqref{eq:edgeHamiltonianSC} can be regarded as two ``copies'' --- ``particle'' and ``hole'', related by PH symmetry --- of the low-energy edge theory defined by Eqs.~\eqref{eq:edgeHamiltonianQH}-\eqref{eq:periodicBCPhi} for quantum Hall phases. The fields $\bar{\varphi}_\sigma$ represent the hole equivalent of $\varphi_\sigma$, in a Bogoliubov de-Gennes (BdG) picture where particles and holes are treated as independent and, hence, internal degrees of freedom are artificially doubled (see Appendix~\ref{sec:bosonizedPictureSC}). Particle and hole fields $\varphi_\sigma$ and $\bar{\varphi}_\sigma$ satisfy the same commutation relations as in Eq.~\eqref{eq:KacMoodyQH1} [and Eq.~\eqref{eq:KacMoodyQH2}, for distinct fields]. The vector $(\varphi_\sigma, \bar{\varphi}_\sigma)^T$ can be regarded as a Nambu spinor. Though $\varphi_\sigma$ and $\bar{\varphi}_\sigma$ are independent in Nambu space, the subspace of physical operators is identified by the ``reality condition''~\footnote{Fermionic particles and holes are created by vertex operators proportional to $(\Psi_\sigma^\text{f})^\dagger = \exp[-i l \varphi_\sigma]$ and $\Psi_\sigma^\text{f} = \exp[i l \varphi_\sigma]$, respectively.}
\begin{equation} \label{eq:realityCondition}
    \bar{\varphi}_\sigma = -\varphi_\sigma.
\end{equation}
By analogy with quantum Hall phases [Eq.~\eqref{eq:edgeHamiltonianQH}], we identify $(\Psi_\sigma^\text{qp})^\dagger = \exp[-i \varphi_\sigma]$ and $(\Psi_\sigma^\text{qh})^\dagger = \exp[-i \bar{\varphi}_\sigma] = \Psi_\sigma^\text{qp}$ as creation operators for Laughlin quasiparticles and quasiholes, respectively.

Under flux variations $\delta\Phi$ and $\delta\chi$, the fields $\varphi_\sigma$ and $\bar{\varphi}_\sigma$ are modified according to Eq.~\eqref{eq:minimalCouplingChiralBosonicFields} --- with $e/l \to -e/l$ for $\bar{\varphi}_\sigma$, in agreement with the fact that $(\Psi_\sigma^\text{qp})^\dagger$ and $(\Psi_\sigma^\text{qh})^\dagger$ carry opposite charges $e/l$ and $-e/l$. In momentum space, 
we obtain, in a similar way as in Eq.~\eqref{eq:commutatorHPsiQp},
\begin{align}
\begin{split} \label{eq:commutatorHPsiQpQhMomSpace}
    \left[ H_\sigma, [(\Psi_\sigma^\text{qp})^\dagger(p)]^s \right] = \sigma v_\sigma \left( p - q_s \delta A_\sigma \right) [(\Psi_\sigma^\text{qp})^\dagger(p)]^s, \\
    \left[ H_\sigma, [(\Psi_\sigma^\text{qh})^\dagger(p)]^s \right] = \sigma v_\sigma \left( p + q_s \delta A_\sigma \right) [(\Psi_\sigma^\text{qh})^\dagger(p)]^s,
\end{split}
\end{align}
where $q_s = (s/l)e$ with $1 \leq s \leq l$, and $p = 2\pi n/L_x$ with integer $n$. These expressions allow us to identify the relevant (PH-symmetric) low-energy edge quasiparticles of the system: superpositions of Laughlin quasiparticles and quasiholes with center-of-mass momentum $2 q_s \delta A_\sigma$, created by operators
\begin{equation} \label{eq:chiralMajoranaModes}
    \gamma_{\sigma, s}^\dagger(p) = \frac{1}{2} \left( [(\Psi_\sigma^\text{qp})^\dagger(p)]^s + [(\Psi_\sigma^\text{qh})^\dagger(p - 2 q_s \delta A_\sigma)]^s \right).
\end{equation}
The corresponding energy dispersion is, as in Eq.~\eqref{eq:anomalousEdgeSpectralFlow},
\begin{align}
    E_{\sigma, s}(p) &= \sigma v_\sigma \left( p - q_s \delta A_\sigma \right) \nonumber \\
    &= \sigma v_\sigma \left[ p - q_s \frac{\delta\Phi + (1 + \sigma) \delta\chi/2}{L_x} \right],\label{eq:anomalousEdgeSpectralFlowSC}
\end{align}
where we use the gauge defined in Eq.~\eqref{eq:gaugeChoice1}, here and in the remaining of this work, for convenience.

Low-energy edge quasiparticles created by $\gamma_{\sigma, s}^\dagger(p)$ are known as chiral Majorana modes (fractional ones, when $l > 1$ and $s < l$)~\cite{Kitaev2006,Qi2010,Qi2011,Lindner2012,Cheng2012,Clarke2013,He2017}. They do not carry any charge, as their constituent Laughlin quasiparticles and quasiholes carry opposite charges. The reality condition in Eq.~\eqref{eq:realityCondition} implies that $(\Psi_\sigma^\text{qp})^\dagger(p) = \Psi_\sigma^\text{qh}(-p)$, and
\begin{align}
    \gamma_{\sigma, s}^\dagger(p) &= \gamma_{\sigma, s}(-p + 2q_s \delta A_\sigma), \\ \label{}
    E_{\sigma, s}(p) &= -E_{\sigma, s}(-p + 2q_s \delta A_\sigma). \label{}
\end{align}
Therefore, modes with energy $E_{\sigma, s}(p) \geq 0$ can be regarded as the only physically distinct degrees of freedom.

Modes with $E_{\sigma, s}(p) = 0$, known as Majorana zero modes~\cite{Read2000,Kitaev2001,Alicea2012}, appear at the edge $\sigma$ with momentum $p_{\sigma, 0} \equiv q_s \delta A_\sigma$ and Hermitian operator $\gamma_{\sigma, s}^\dagger(p_{0, \sigma}) = \gamma_{\sigma, s}(p_{0, \sigma})$ provided that $p_{\sigma, 0}$ is an allowed momentum value, i.e., if and only if
\begin{equation} \label{eq:zeroModeCondition}
    L_x p_{\sigma, 0} = q_s \left[ \delta\Phi + (1 + \sigma) \delta\chi/2 \right] = 2\pi n,
\end{equation}
for some integer $n$. Majorana zero modes therefore appear at both edges when $\delta\Phi = \delta\chi = 0$ (at $p_{0, \sigma} = 0$, for any $1 \leq s \leq l$)~\footnote{On a Corbino disk with $\delta\chi = 0$, Majorana zero modes would require a shift of $\Phi$ by an odd number of superconducting flux quanta, $\Phi \to \Phi + \pi$, to compensate for the fact that the extrinsic curvature of a cylinder is lost upon deformation to a Corbino disk (see, e.g., Ref.~\cite{Bardyn2012}).}, and remain present for $\delta\Phi = (2\pi l) m$ with integer $m$ (corresponding to an even number $m$ of superconducting flux quanta $\pi$, when $l = 1$). Flux variations $\delta\chi$ only affect modes at the right edge ($\sigma = +1$). When $\delta\Phi = (2\pi l) m$, they preserve the Majorana zero mode at the right edge provided that $\delta\chi = (2\pi l) n$ with integer $n$ too. Particle-hole symmetry ensures that zero modes always come in pairs~\footnote{Majorana zero modes are protected by an index theorem and particle-hole symmetry~\cite{Tewari2007,Cheng2010}, which imply that they can only appear or be lifted in pairs.}. Therefore, any zero mode that disappears from the right edge due to $\delta\chi$ must appear in the bulk where $\delta\chi$ is inserted. We discuss such a situation below and in Appendix~\ref{sec:tightBindingSC}.

\begin{figure}[t]
\begin{center}
    \includegraphics[width=\columnwidth]{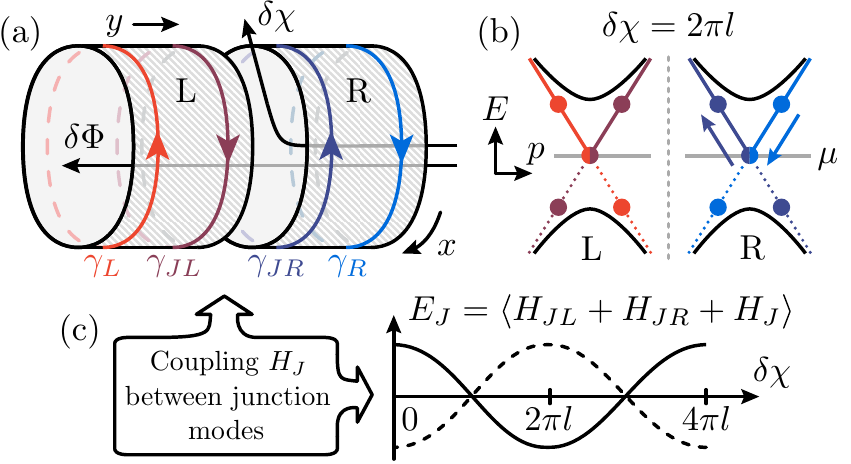}
    \caption{Effects of bulk flux quanta $\delta\chi$ in 2D topological superconductors. \textbf{(a)}: To insert $\delta\chi$ through the superconducting bulk, we split the system into two parts, ``left'' (L) and ``right'' (R). When these parts are fully disconnected, a pair of chiral Majorana modes $\gamma_{L, s}(p)$ and $\gamma_{R, s}(p)$ (shown with simplified notations) appears at the edges, with another pair $\gamma_{JL, s}(p)$ and $\gamma_{JR, s}(p)$ at the junction. \textbf{(b)}: Accordingly, the spectra of left and right parts exhibit a pair of counterpropagating chiral Majorana modes centered around momentum $p = 0$ and $p = q_s \delta\chi/L_x$, respectively (setting $\delta\Phi = 0$ as in the text), where $L_x$ is the system's length in the $x$ direction. Shown here is the integer case $l = 1$ (see text), where dots represent non-fractionalized single-particle chiral Majorana states. Arrows illustrate the spectral flow induced by $\delta\chi = 2\pi l$, which only acts on the right part of the system. \textbf{(c)}: When introducing a weak coupling $H_J$ between junction modes (with low-energy Hamiltonian $H_{JL}$ and $H_{JR}$, respectively), the change in energy $E_J = \av{H_{JL} + H_{JR} + H_J}$ exhibits a $\delta\chi = 4\pi l$ periodicity (assuming that parity is conserved; see text), corresponding to a fractional Josephson current $I_J = \partial E_J/\partial \chi$ with the same periodicity. The energy levels sketched here are the two fermionic modes resulting from the hybridization of the junction Majorana modes, for $l = 1$.}
    \label{fig:josephsonJunction}
\end{center}
\end{figure}

We now examine the effects of bulk flux variations $\delta\chi$ more broadly, setting $\delta\Phi = 0$, without loss of generality. To insert $\delta\chi$ in the superconducting bulk, we consider a slightly modified system where superconductivity is weak or absent in a narrow annular region of the bulk (see Fig.~\ref{fig:josephsonJunction}a). This setup can be regarded as a Josephson junction or weak link between two cylindrical topological superconductors, ``left'' (L) and ``right'' (R). Flux variations $\delta\Phi$ thread both superconductors, whereas $\delta\chi$ threads the right one only. The details of the junction are essentially irrelevant for our purposes (an explicit model can be found in Appendix~\ref{sec:tightBindingSC}, for $l = 1$).

We first examine the situation where left and right superconducting parts of the bulk are completely disconnected. In that case, each part exhibits a pair of chiral Majorana modes: one at an edge of the whole system, and one at the junction. We denote the edge modes as $\gamma_{L, s}(p)$ and $\gamma_{R, s}(p)$, as in Eq.~\eqref{eq:chiralMajoranaModes}, and the junction (bulk) modes as $\gamma_{JL, s}(p)$ and $\gamma_{JR, s}(p)$. Since the spectra of chiral Majorana modes and Laughlin quasiparticles have the same form [compare Eqs.~\eqref{eq:anomalousEdgeSpectralFlow} and~\eqref{eq:anomalousEdgeSpectralFlowSC}], flux variations induce the same anomalous spectral flow, here in Nambu space, as in the quantum Hall phases examined above. In particular, the spectrum of low-energy edge and junction modes is invariant under variations $\delta\Phi = (2\pi l) m$ and $\delta\chi = (2\pi l) n$ with integer $m, n$. 

From the viewpoint of the edge modes, $\delta\chi = 2\pi l$ pumps exactly one chiral Majorana \emph{fermion} $\gamma_{R, l}(p)$ (with $s = l$) across the zero-energy (Fermi) level, from the right edge into the bulk (see Fig.~\ref{fig:josephsonJunction}b). This Majorana fermion represents a superposition of a quasiparticle and a quasihole with unit charge [$q_l = e$ in Eq.~\eqref{eq:chiralMajoranaModes}]. When it crosses zero energy, the energy of these constituents changes sign [Eq.~\eqref{eq:commutatorHPsiQpQhMomSpace}], and one can distinguish two scenarios: (i) If the system is connected to an external reservoir of charges, the $\mathbb{Z}_2$ fermion-number parity of the ground state changes. (ii) If parity is conserved, instead, $\delta\chi = 2\pi l$ leads to an excited state, and $\delta\chi = 4\pi l$ is required for the system to come back to its initial ground state and parity. In summary, $\delta\chi = (2\pi l) n$ induces real [case (i)] or apparent [case (ii)] parity switches, in a similar way as in quantum Hall examples. Here we focus on the case (ii) with parity conservation. The key difference with the quantum Hall case where the fermion number is conserved is twofold: First, the bulk is invariant under double parity switches. Second, the edge-mode velocities $v_\sigma$ in Eq.~\eqref{eq:edgeHamiltonianSC} are not modified~\footnote{Provided that $\delta\chi$ is introduced sufficiently deep into the bulk, where modes have an exponentially small overlap with edge modes; see Appendix~\ref{sec:tightBindingSC}.}. In topological superconductors, $\delta\chi = (4\pi l)n$ (with integer $n$) thus represents a bona fide topological pump.

We now switch on the coupling between left and right parts of the system, and examine the behavior of the latter at the junction where $\delta\chi$ is inserted. From the viewpoint of junction modes, $\delta\chi$ modifies the energy $E_J = \av{H_{JL} + H_{JR} + H_J}$ arising from the weak coupling $H_J$ between $\gamma_{JL, s}(p)$ and $\gamma_{JR, s}(p)$ (described by low-energy effective Hamiltonians $H_{JL}$ and $H_{JR}$, respectively, where $\av{...}$ denotes the ground-state expectation value). This flux-dependent energy modification gives rise to a dc Josephson supercurrent~\cite{Josephson1964} between left and right superconductors,
\begin{equation} \label{eq:josephsonCurrent}
    I_J = \frac{\partial E_J}{\partial \chi}.
\end{equation}
The junction modes $\gamma_{JL, s}(p)$ and $\gamma_{JR, s}(p)$ experience the same anomalous spectral flow as the edge modes $\gamma_{L, s}(p)$ and $\gamma_{R, s}(p)$. In particular, $\delta\chi = 2\pi l$ pumps one chiral Majorana fermion in mode $\gamma_{JR, l}(p)$ across the Fermi level, leading to an excited state with an apparent parity switch. Assuming that parity is conserved, the energy $E_J$ and current $I_J$ are periodic under $\delta\chi = (4\pi l) n$ with integer $n$, for an arbitrary constant coupling $H_J$ between junction modes. In other words, the topological pump $\delta\chi = (4\pi l)n$ identified above from the behavior of edge modes manifests itself as a $4\pi l$ periodic Josephson current in the bulk where $\delta\chi$ is inserted (see Fig.~\ref{fig:josephsonJunction}c). We thus find a direct connection between topological pumping and the so-called fractional Josephson effects identified in other settings focusing on Majorana zero modes~\cite{Kitaev2001,Fu2008,Fu2009} and their fractional analogs ($l > 1$)~\cite{Lindner2012,Cheng2012,Clarke2013}.

We remark that $\delta\chi$ controls the superconducting phase difference across the junction. Indeed, generic fermionic fields $\Psi^\text{f}(x)$ on either side of the junction transform as $\Psi^\text{f}(x) \to \Psi^\text{f}(x) \exp[-i \int_0^x dx' \delta A_\sigma(x')]$ under flux variations [recall Eq.~\eqref{eq:minimalCouplingChiralBosonicFields}]. In our chosen gauge [Eq.~\eqref{eq:gaugeChoice1}], $\delta\chi$ induces a phase $\exp[-i (\delta\chi/L_x)x]$ for fermionic fields in the right superconductor, corresponding to a phase change $\Delta(x,x') \to \Delta(x,x') \exp[i (\delta\chi/L_x)(x + x')]$ for the superconducting order parameter $\Delta(x,x')$ of the latter (see Appendix~\ref{sec:tightBindingSC} for an explicit example). On ``average'', the superconducting phase difference across the junction (with $x' = x$) is therefore $(1/L) \int_0^L dx (2\delta\chi/L_x) x = \delta\chi$. Note that fluxes $\Phi$ and $\chi$ take quantized values, in practice, corresponding to integer multiples of the superconducting flux quantum $\pi$ (i.e., $\delta\Phi = \pi m$ and $\delta\chi = \pi n$ with integer $m, n$)~\cite{Byers1961}. This flux quantization is required, in particular, for the center-of-mass momentum $2\delta\chi/L_x$ of Cooper pairs to be commensurate with allowed momenta.

We remark that the identification of $\delta\chi = (4\pi l)n$ as a low-energy topological pump is independent of the way $\delta\chi$ is threaded through the bulk~\footnote{Provided that $\delta\chi$ is inserted deep into the bulk, in a region with negligible overlap with the edge modes.}. In particular, $\delta\chi$ could be inserted in the form of vortices. In that case, $\delta\chi = 2\pi l$ would correspond to the introduction of a vortex carrying a bound Majorana zero mode (see, e.g., Refs.~\cite{Read2000,Alicea2012}), and $\delta\chi = 4\pi l$ would correspond to the introduction and subsequent ``fusion'' of two such vortices, known to leave the system invariant~\cite{Nayak2008,Alicea2012}.


\section{Conclusions}
\label{sec:conclusions}

We have shown that the insertion of flux quanta $\delta\chi$ through the bulk of gapped phases with topological gapless edge states provides a generic knob for robust physical phenomena. The robustness of these effects can be traced to the direct connection between $\delta\chi$ and topological (Thouless) pumping, which originates from anomalous low-energy (edge) spectral flows. We have demonstrated this generic connection in two paradigmatic types of noninteracting (integer) and interacting (fractional) topological phases: 2D quantum Hall systems and topological superconductors.

In the quantum Hall phases examined here (supporting elementary quasiparticle excitations with charge $e/l$), we have shown that flux variations $\delta\chi = 4\pi l$ result in the injection of two fermions (charges) from the edges into the bulk, which can be observed, e.g., via persistent currents in a mesoscopic system~\cite{Filippone2017}. In superconducting analogs of these phases, in contrast, we have shown that $\delta\chi = 4\pi l$ induces an apparent double parity switch, which can be seen, e.g., in Josephson currents.

Although parity switches in persistent currents, fractional Josephson currents, and other effects of bulk magnetic fluxes have been explored in a variety of settings (see, e.g., Refs.~\cite{Kitaev2001,Fu2008,Fu2009,Lindner2012,Cheng2012,Clarke2013,Filippone2017}), our work identifies the concept of bulk pumping as a general framework to derive and understand such seemingly distinct effects in a systematic way. It will be interesting to explore the robust effects of bulk pumping in more exotic types of topological phases with protected edge or higher-dimensional surface states, such as topological (crystalline) insulators~\cite{Hasan2010,Fu2011,Hsieh2012,Po2012}, Weyl and Dirac semimetals~\cite{Hasan2017,Yan2017,Armitage2018}, or simulated four-dimensional quantum Hall systems~\cite{Zilberberg2018,Lohse2018}. In the same vein as Laughlin's conventional topological pumping, bulk pumping provides a practical probe of gapped topological phases with surface states, irrespective of the presence of disorder and interactions.


\emph{Acknowledgements}--- We thank Dmitri Abanin, Nigel R. Cooper, Thierry Jolicoeur, Ivan Protopopov, Steven H. Simon, and Luka Trifunovic for useful discussions, and gratefully acknowledge support by the Swiss National Science Foundation under Division II.

\appendix

\section*{Appendices}

\section{Bulk pumping in an explicit tight-binding model for integer ($l = 1$) quantum Hall phases}
\label{sec:tightBindingQH}

In this first Appendix, we revisit the example of integer ($l = 1$) quantum Hall phases in an explicit noninteracting tight-binding model. Our goal is to detail the effects of flux variations $\delta\Phi$ and $\delta\chi$ in a minimal concrete setting including bulk and edge modes.

\begin{figure}[t]
\begin{center}
    \includegraphics[width=\columnwidth]{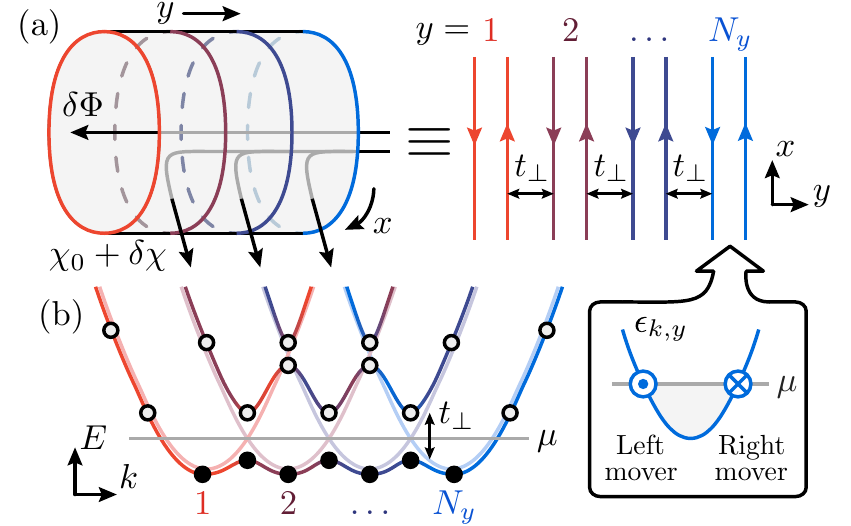} 
    \caption{Coupled-wire description of the system in a noninteracting integer quantum Hall phase with filling factor $\nu = 1$. \textbf{(a)}: Generic setup consisting of $N_y$ 1D wires (red to blue) wrapped around a cylinder threaded by Aharonov-Bohm and transverse fluxes $\Phi$ and $\chi$. A constant background flux $\chi = \chi_0$ is present, tuned with the Fermi level $\mu$ so that $\nu = 1$. Low-energy degrees of freedom are illustrated on the right: Each wire with index $y$ and energy dispersion $\epsilon_{k,y}$ exhibits a pair of left- and right-moving modes at the Fermi level (fermionic modes with unit charge). Neighboring wires are coupled by some tunneling $\delta_\perp$ [Eq.~\eqref{appEq:interwireCouplingQH}] so that right movers in wire $y$ couple to left movers in wire $y+1$, leaving a pair of uncoupled modes at the edges. \textbf{(b)}: Resulting (schematic) single-particle band structure: Bands of individual wires (faint colors) are shifted with respect to each other due to $\chi$ [Eq.~\eqref{appEq:wireHamMomSpaceQH}]. They resonantly couple at the Fermi level, creating a gapped phase with counterpropagating gapless edge modes.}
    \label{fig:coupledWireIQH}
\end{center}
\end{figure}

Following the approach of Refs.~\cite{Kane2002,Teo2014}, we construct the quantum Hall phase of interest in an array of coupled 1D wires. Specifically, we consider a set of $N_y$ identical parallel wires wrapped around a cylinder (Fig.~\ref{fig:coupledWireIQH}a), with periodic boundary conditions in the $x$ direction, and indices $y = 1, \ldots, N_y$ corresponding to wire positions in the $y$ direction (with unit inter-wire spacing). We assume that each wire can be modeled as a translation-invariant lattice system 
of noninteracting spinless fermions with unit charge, with one site per unit cell and a total of $N_x$ cells. We set the lattice spacing along wires to unity, so that wires have a length $L_x = N_x$.

As in the main text, the system is exposed to an Aharonov-Bohm flux $\Phi$ and a bulk transverse flux $\chi$. Assuming that $\chi$ is uniform, the total flux threading each wire (or ring) is $\Phi + (y-1)\chi/(N_y-1)$. In a Landau gauge consistent with Eq.~\eqref{eq:gaugeChoice1} in the main text, this is described by a $U(1)$ gauge field $\mathbf{A}(x,y) = \mathbf{A}(y)$ with $x$ component
\begin{equation} \label{appEq:gaugeField}
    A(y) = \frac{1}{N_x} \left[ \Phi + (y-1) \frac{\chi}{N_y-1} \right],
\end{equation}
and vanishing $y$ component. The system's charges minimally couple to this field, leading to momentum shifts
\begin{equation} \label{appEq:momentumShifts}
    k \to k - A(y)
\end{equation}
in individual wires, where $k$ is the crystal momentum in the $x$ direction ($k = 2\pi n/N_x$ with $n = 0, 1, \ldots, N_x-1$). In terms of the creation operators $c^\dagger_{x,y}$ for fermions on site $x$ of wire $y$ (satisfying periodic boundary conditions $c_{x+N_x,y} = c_{x,y}$), these shifts are described by the replacement $c_{x,y} \to e^{iA(y)x} c_{x,y}$ in the Hamiltonian. The Hamiltonian of individual wires then takes the generic form
\begin{equation} \label{appEq:wireHamMomSpaceQH}
    H_y = \sum_{k} \xi_{k - A(y)} c^\dagger_{k,y} c_{k,y},
\end{equation}
where $\xi_{k - A(y)}$ is the momentum-shifted energy dispersion (band) of each wire $y$. We assume that $\xi_{k - A(y)}$ is such that wires exhibit a pair of left- and right-moving chiral fermionic modes at the Fermi level $\mu$, as shown in the inset of Fig.~\ref{fig:coupledWireIQH}a. For example, $\xi_{k - A(y)} \equiv \mu + \epsilon_{k,y} = \mu - t_\parallel \cos[k - A(y)]$, for a simple tunnel coupling between nearest-neighboring sites with strength $-t_\parallel/2$.

To generate the integer quantum Hall phase of interest, we start with a uniform ``background'' transverse flux $\chi = \chi_0$, which induces a relative momentum shift $\Delta k = \chi_0/[N_x(N_y-1)]$ between energy bands of neighboring wires (Fig.~\ref{fig:coupledWireIQH}b). We tune $\Delta k$ and the Fermi momentum $k_F$ of uncoupled wires so that the filling factor is $\nu = 2k_F/\Delta k = 1$ [this requires a macroscopic flux $\chi_0 \geq 4\pi (N_y-1)$]. In that case, bands of uncoupled wires cross at the Fermi level where wires exhibit left- and right-moving modes, such that right movers in wire $y$ are resonant with left movers in wire $y + 1$. We open a topological gap at these crossings by introducing the following tunnel coupling between neighboring wires:
\begin{equation} \label{appEq:interwireCouplingQH}
    H_{y, y+1} = -\frac{t_\perp}{2} \sum_{k} \left( c^\dagger_{k,y+1} c_{k,y} + h.c. \right),
\end{equation}
where $t_\perp > 0$. The resulting spectrum is illustrated in Fig.~\ref{fig:coupledWireIQH}b: As desired, a pair of counterpropagating chiral fermionic modes with velocity $v_\sigma \approx v_F$ ($v_F$ being the Fermi velocity of uncoupled wires) remains ungapped at the edges: the left- and right-moving modes originating from wires $y = 1$ and $y = N_y$. These two modes are exponentially localized around these edge wires~\footnote{With localization length controlled by $1/t_\perp$.}, and are topologically protected against quasilocal perturbations by their spatial separation. They govern the low-energy physics of the system, described by Eqs.~\eqref{eq:edgeHamiltonianQH}-\eqref{eq:periodicBCPhi} in the main text (in bosonized form, setting $l = 1$).

Under flux variations $\delta\Phi$ and $\delta\chi$, gapless edge modes experience the anomalous spectral flow~\cite{Wen2007,Bernevig2013} discussed in the main text (with $l = 1$, here): Laughlin's pump~\cite{Laughlin1981,Halperin1982} corresponds to the insertion of an Aharonov-Bohm flux $\delta\Phi = 2\pi$ (one flux quantum), which pumps exactly one state from the right edge into the bulk, below the Fermi level $\mu$, and exactly one state from the bulk into the left edge, above $\mu$. The net anomaly outflow from the edges into the bulk vanishes. Bulk flux quanta $\delta\chi = 2\pi$, in contrast, pump only one state from the right edge into the bulk, below the Fermi level $\mu$ (in agreement with Fig.~\ref{fig:setup}b of the main text). The bulk must change in order to absorb the corresponding net anomaly outflow from the edges. Indeed, flux variations $\delta\chi$ modify the spacing between bands of uncoupled wires by $\delta\chi/[N_x(N_y-1)]$ [see Eq.~\eqref{appEq:gaugeField}], leading to spectral modifications of order $\mathcal{O}(\delta \chi/S)$, where $S = N_x N_y$ is the surface area of the system. Since edge modes are off-resonantly coupled to bulk modes by the inter-wire couplings, their velocity $v_\sigma \approx v_F$ is modified by a small correction $\mathcal{O}(\delta \chi/S)$.

\section{Bulk pumping in an explicit tight-binding model for integer ($l = 1$) topological superconducting phases}
\label{sec:tightBindingSC}

We now detail the effects of flux variations $\delta\Phi$ and $\delta\chi$ in an explicit noninteracting tight-binding model for integer ($l = 1$) topological superconducting phases. 2D topological superconductors can be constructed from coupled 1D wires in a similar way as in Appendix~\ref{sec:tightBindingQH}~\cite{Mong2014,Neupert2014,Alicea2015,Huang2016,Sagi2017}. As for quantum Hall phases, the desired topological phase can be obtained by coupling right- and left-moving modes in neighboring wires $y$ and $y+1$ in such a way that gapless modes remain at the edges only. Here, however, right and left movers can be made resonant without background transverse flux $\chi_0$, as detailed below.

\begin{figure}[t]
\begin{center}
    \includegraphics[width=\columnwidth]{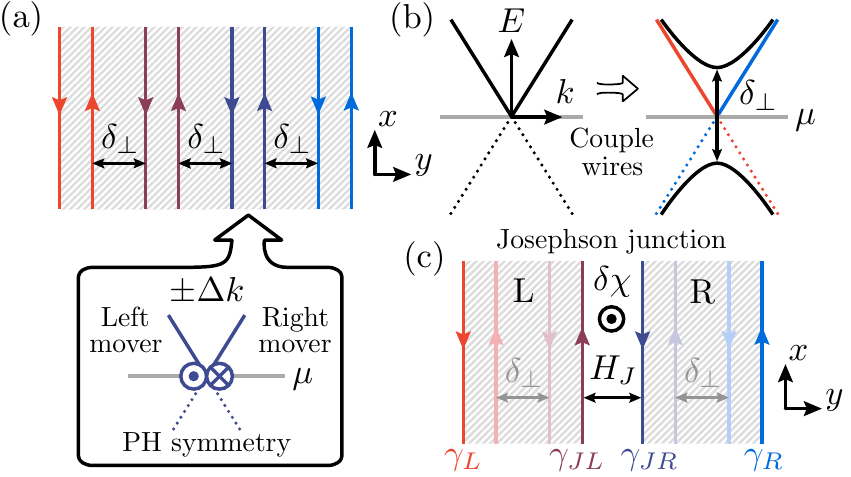} 
    \caption{Coupled-wire description of the system in a noninteracting (integer) topological superconducting phase. \textbf{(a)}: Same setup as in Fig.~\ref{fig:coupledWireIQH}a, with two modifications: (i) no background flux $\chi_0$, and (ii) superconducting pairings induced, e.g., by an underlying superconductor (grey striped background). Each wire corresponds to a gapless Kitaev chain~\cite{Kitaev2001}, i.e., to a 1D superconductor with a pair of left- and right-moving modes crossing at the Fermi level $\mu$ [Eq.~\eqref{appEq:linearizedWireHamMomSpaceSC}]. Due to particle-hole (PH) symmetry, each mode represents a superposition of quasiparticles and quasiholes (fermionic ones, with unit charge). Neighboring wires are coupled by a combination $\delta_\perp$ of tunneling and pairing [Eq.~\eqref{appEq:interwireCouplingSC}] so that right movers in wire $y$ couple to left movers in wire $y+1$, leaving a pair of uncoupled modes at the edges. \textbf{(b)}: Resulting (schematic) single-particle band structure: Bands of individual wires (shown in black) resonantly couple, leading to a gapped phase with PH-symmetric counterpropagating gapless edge modes (red and blue). \textbf{(c)}: Josephson junction created by the insertion of bulk flux quanta $\delta\chi$ (see text).}
    \label{fig:coupledWireSC}
\end{center}
\end{figure}

We start from the same coupled-wire array as in Appendix~\ref{sec:tightBindingQH}, with $\Phi = 0$, without loss of generality, and with $\chi_0 = 0$. We then add superconducting pairings induced, e.g., by proximity coupling to a superconductor (see Fig.~\ref{fig:coupledWireSC}a). In the absence of flux variations $\delta\Phi$ and $\delta\chi$ (i.e., for $\Phi = \chi = 0$), the Hamiltonian of individual wires takes the standard particle-hole (PH) symmetric Bogoliubov-de Gennes (BdG) form
\begin{align}
    H^{(0)}_y &= \frac{1}{2} \sum_{k} \Psi^\dagger_{k,y} \mathcal{H}_{k} \Psi_{k,y} + E_0, \\
    \mathcal{H}_{k} &= \left( \begin{array}{cc}
         \xi_{k} & \Delta_{k} \\
         \Delta^*_{k} & -\xi_{-k}
    \end{array} \right), \label{appEq:BdGHamMatrix}
\end{align}
where $\Psi^\dagger_{k,y} = (c^\dagger_{k,y}, c_{-k,y})$, $\xi_{k}$ is the energy dispersion of individual wires, and $E_0 = (1/2) \sum_{k} \xi_{k}$ is an energy shift which we set to zero~\footnote{This zero-energy level corresponds to the Fermi level.}. Pairings are described by $\Delta_{k}$.

As in the main text, we consider the effects of transverse flux quanta $\delta\chi$ inserted deep into the bulk, say, between wires $Y$ and $Y+1$ in the middle of the latter (choosing $Y = N_y/2$ for even $N_y$). The resulting system can be regarded as two parts [left (L) and right (R), as in Fig.~\ref{fig:josephsonJunction}a of the main text] coupled to distinct uniform gauge fields
\begin{align} \label{appEq:gaugeFieldLeftRightSCs}
    A(y) = \begin{cases}
        0, \quad & y \leq Y, \\
        \frac{\delta\chi}{N_x}, \quad & y \geq Y+1,
    \end{cases}
\end{align}
where we have used the same (Landau) gauge as in Eq.~\eqref{eq:gaugeChoice1} of the main text. The flux $\delta\chi$ threads the right part of the system alone, inducing a uniform momentum shift $k \to k - \delta\chi/N_x$ in the latter. As in Appendix~\ref{sec:tightBindingQH}, the minimal coupling of system's charges to $\delta\chi$ is described by the replacement $c_{x,y} \to e^{i A(y)x} c_{x,y}$ in the Hamiltonian. 
The Hamiltonian of uncoupled wires becomes
\begin{equation} \label{appEq:wireHamMomSpaceSC}
    H_y = \frac{1}{2} \sum_{k} \tilde{\Psi}^\dagger_{k,y} \mathcal{H}_{k - A(y)} \tilde{\Psi}_{k,y},
\end{equation}
where $\tilde{\Psi}^\dagger_{k,y} = (c^\dagger_{k,y}, c_{-k + 2A(y),y})$, with $A(y)$ given by Eq.~\eqref{appEq:gaugeFieldLeftRightSCs}. The spectrum of individual uncoupled wires $y$ is determined by $\mathcal{H}_{k - A(y)}$: It is centered around $k = A(y)$ (i.e., around $k = 0$ and $k = \delta\chi/N_x$ for wires in the left and right parts of the system, respectively), with eigenenergies satisfying $E_{k - A(y)} = -E_{-k + A(y)}$, due to PH symmetry~\footnote{The relevant PH symmetry is given by $C \mathcal{H}_{k - A(y)} C^{-1} = - \mathcal{H}_{-k + A(y)}$ with $C = \sigma_x K$, where $K$ is the antiunitary operator describing time-reversal symmetry (or complex conjugation in position space).}. Pairing occurs between fermionic quasiparticles and quasiholes with momentum $k$ and $-k + 2A(y)$, respectively, corresponding to Cooper pairs with center-of-mass momentum $2A(y)$~\footnote{In accordance with the charge $2e$ of Cooper pairs.}. As we consider spinless fermions, the pairing function $\Delta_{k}$ is odd under $k \to -k$. In particular, $\Delta_{k - A(y)}$ vanishes at $k = A(y)$ in wire $y$.

To generate a topological phase in each part of the system (left and right), we follow a similar strategy as for quantum Hall phases in Appendix~\ref{sec:tightBindingQH}: We try to reach a situation where individual wires support a pair of left- and right-moving modes, and introduce a suitable inter-wire coupling to make right movers in wire $y$ couple to left movers in wire $y + 1$, to open a gap in the bulk while leaving gapless edge modes. Due to PH symmetry, wires can only exhibit chiral modes if they are gapless. We thus start from gapless uncoupled wires, tuning the chemical potential so that $\xi_{k - A(y)} = 0$ at $k = A(y)$~\footnote{Although this fine tuning makes our construction more transparent, it is not required. The chemical potential need only be tuned within a range of the order of the topological gap.}. Assuming that $\Delta_{k - A(y)}$ vanishes linearly at $k = A(y)$ (as in $p$-wave superconductors~\cite{Read2000,Kitaev2001,Alicea2012}), the low-energy physics of uncoupled wires is given by Eq.~\eqref{appEq:wireHamMomSpaceSC} with
\begin{equation} \label{appEq:linearizedWireHamMomSpaceSC}
    \left. \mathcal{H}_{k - A(y)} \right|_{k \approx A(y)} \approx \Delta [k - A(y)] \sigma_y,
\end{equation}
where $\sigma_{y}$ denotes the standard Pauli matrix. As desired, wires in each part of the system exhibit a pair of right- and left-moving modes (corresponding to the eigenstates of $\sigma_y$ with eigenvalues $\pm 1$) with velocity $v_\sigma = \Delta$. Explicitly, each wire supports low-energy chiral Majorana fermionic modes given by
\begin{equation} \label{appEq:leftRightModesSC}
    \left( \begin{array}{c}
         \gamma^+_{k,y} \\
         \gamma^-_{k,y}
    \end{array} \right)
    = \frac{1}{\sqrt{2}}
    \left( \begin{array}{cc}
         e^{-i\pi/4} & e^{i\pi/4} \\
         e^{i\pi/4} & e^{-i\pi/4}
    \end{array} \right)
    \tilde{\Psi}_{k,y}.
\end{equation}
Consequently, in each part of the system, the desired inter-wire couplings between right movers ($+$) in wire $y$ and left movers ($-$) in wire $y + 1$ read
\begin{align} \label{appEq:interwireCouplingSC}
    H_{y,y+1} &= i\frac{\delta_\perp}{2} \sum_{k} (\gamma^-_{k,y+1})^\dagger \gamma^+_{k,y} + h.c. \nonumber \\
    &= \frac{1}{2} \sum_{k} \tilde{\Psi}^\dagger_{k,y+1} \left( -\frac{t_\perp}{2} \sigma_z + i \frac{\Delta_\perp}{2} \sigma_x \right) \tilde{\Psi}_{k,y} + h.c.,
\end{align}
where we recall that $\tilde{\Psi}^\dagger_{k,y} = (c^\dagger_{k,y}, c_{-k + 2A(y),y})$, with $A(y)$ given by Eq.~\eqref{appEq:gaugeFieldLeftRightSCs}. These inter-wire couplings represent a combination of tunnel coupling and superconducting pairing between nearest-neighboring wires in each part of the system, with equal amplitude $t_\perp = \Delta_\perp \equiv \delta_\perp$. 

By construction, the set of inter-wire couplings $H_{y,y+1}$ (for all $y = 1, \ldots, N_y$ except $y = Y$) gap out all low-energy modes except for one pair of counterpropagating edge modes in each part of the system: the left- and right-moving modes of wires $y = 1$ and $y = Y$, $\gamma^-_{k,0} \equiv \gamma_L$ and $\gamma^+_{k,Y} \equiv \gamma_{JL}$, and the left- and right-moving modes of wires $y = Y+1$ and $y = N_y$, $\gamma^-_{k,Y+1} \equiv \gamma_{JR}$ and $\gamma^+_{k,N_y} \equiv \gamma_{R}$ (see Fig.~\ref{fig:coupledWireSC}c, and Fig.~\ref{fig:josephsonJunction}a of the main text). We call $\gamma_{JL}$ and $\gamma_{JR}$ the junction modes. These four integer (non-fractionalized, $l = 1$) chiral Majorana fermionic modes with velocity $v_\sigma = \Delta$ govern the low-energy physics of the system when left and right parts are decoupled ($H_J = 0$ in Fig.~\ref{fig:coupledWireSC}c). Each mode is described by an effective Hamiltonian of the form of Eq.~\eqref{eq:edgeHamiltonianSC} in the main text (in bosonized form, with $l = 1$), with Eq.~\eqref{appEq:leftRightModesSC} providing the analog of Eq.~\eqref{eq:chiralMajoranaModes}, for $l = 1$.

As argued in the main text, the above low-energy modes are superpositions of chiral fermionic quasiparticles and quasiholes as found in the integer quantum Hall case [$c_{k,y}$ and $c^\dagger_{-k + 2A(y),y}$ in Eq.~\eqref{appEq:leftRightModesSC}]. Consequently, their spectral flow behaves in a similar way: Each bulk flux quantum $\delta\chi = 2\pi$ pumps exactly one chiral Majorana fermionic mode across the Fermi level $\mu$, from the right edge into the bulk (see Fig.~\ref{fig:josephsonJunction}b of the main text). If the system is connected to an external reservoir of particles, the parity of the ground state changes in the process. Here, however, $\delta\chi = 2\pi$ leads to an excited state with an apparent parity switch only, and $\delta\chi = 4\pi$ is required to come back to the original ground state. Low-energy observables are therefore typically $4\pi$ periodic in $\delta\chi$.

In contrast to what we found for quantum Hall phases in Appendix~\ref{sec:tightBindingQH}, $\delta\chi$ does not modify, here, the velocity $v_\sigma$ ($= \Delta$) of the edge modes. The bulk pump $\delta\chi = 4\pi$ leaves the low-energy theory described by $\gamma_L$, $\gamma_{JL}$, $\gamma_{JR}$ and $\gamma_{R}$ invariant and, hence, represents a bona fide low-energy topological pump. Intuitively, this can be understood as follows: $\delta\chi = 4\pi$ pumps two (chiral Majorana fermionic) states below the zero-energy (Fermi) level, as in Fig.~\ref{fig:setup}b of the main text. Due to PH symmetry, this corresponds to an apparent double parity switch, or to the injection of an additional Cooper pair into the bulk.

\begin{figure}[t]
\begin{center}
    \includegraphics[width=\columnwidth]{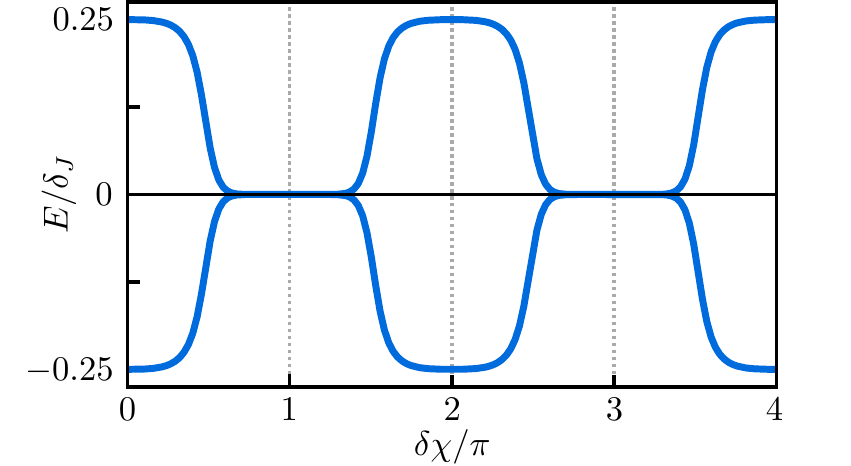}
    \caption{Low-energy BdG spectrum of the noninteracting (integer) topological superconductor described by Eqs.~\eqref{appEq:wireHamMomSpaceSC} and~\eqref{appEq:interwireCouplingSC}, with flux $\delta\chi$ inserted between wires $Y$ and $Y+1$ where superconductivity is absent. The resulting Josephson junction is governed by the weak coupling $H_J$ between these two wires [chosen here as in Eq.~\eqref{appEq:junctionCoupling}], which couples the two chiral Majorana fermionic modes appearing at the junction, when $H_J = 0$ ($\gamma_{JL}$ and $\gamma_{JR}$ in Fig.~\ref{fig:coupledWireSC}). The low-energy spectrum visible here shows the two modes arising from the hybridization of $\gamma_{JL}$ and $\gamma_{JR}$. All other modes are gapped, appearing at higher energies of the order of the bulk inter-wire coupling strength $\delta_\perp$ [Eq.~\eqref{appEq:interwireCouplingSC}]. For clarity, low-energy edge modes ($\gamma_L$ and $\gamma_{R}$ in Fig.~\ref{fig:coupledWireSC}) are also gapped out by a direct coupling of the form of Eq.~\eqref{appEq:interwireCouplingSC}. For even $\delta\chi/\pi$, the pair of Majorana zero modes that appears at the junction is gapped by $H_J$. For odd $\delta\chi/\pi$, in contrast, a single Majorana zero mode is present at the junction, and, hence, the effect of $H_J$ is exponentially small. The above plot shows the actual spectrum corresponding, for $l=1$, to the schematic spectrum in Fig.~\ref{fig:josephsonJunction}c. In general, energy levels cross the Fermi level at odd $\delta\chi/\pi$ in an exponential way. The energy-dispersion and pairing functions in Eq.~\eqref{appEq:BdGHamMatrix} are chosen here as $\xi_k = \mu - t_\parallel \cos(k)$ and $\Delta_k = i \Delta \sin(k)$ with $\mu = t_\perp$, such that decoupled wires correspond to gapless Kitaev chains~\cite{Kitaev2001}.}.
    \label{fig:josephsonEffect}
\end{center}
\end{figure}

As discussed in the main text, the $4\pi$ periodicity of low-energy observables can be observed, e.g., by measuring the Josephson current flowing at the junction between left and right parts of the system, when the latter are weakly coupled with Hamiltonian $H_J$ (Fig.~\ref{fig:coupledWireSC}c). This current is proportional to the energy change $I_J = \partial_\chi \av{H_{JL} + H_{JR} + H_J}$ induced by $\delta\chi$, where $H_{JL}$ and $H_{JR}$ denote the low-energy effective Hamiltonians describing the junction modes $\gamma_{JL}$ and $\gamma_{JR}$, and $\av{...}$ denotes the ground-state expectation value. The most direct coupling between junction modes reads
\begin{align} \label{appEq:directJunctionCoupling}
    H^\text{direct}_J &\equiv H_{Y,Y+1} = i\frac{\delta_J}{2} \sum_{k} (\gamma^-_{k,Y+1})^\dagger \gamma^+_{k,Y} + h.c. \nonumber \\
    &= \frac{1}{2} \sum_{k} \tilde{\Psi}^\dagger_{k,Y+1} \left( -\frac{t_J}{2} \sigma_z + i \frac{\Delta_J}{2} \sigma_x \right) \tilde{\Psi}_{k,Y} + h.c.,
\end{align}
as in the rest of the bulk [Eq.~\eqref{appEq:interwireCouplingSC}], with amplitude $\delta_J = t_J = \Delta_J$. Since superconductivity is suppressed at the junction where the flux $\delta\chi$ is inserted (i.e., $\Delta_J \approx 0$), however, a more natural coupling is
\begin{align} 
    H_J &= \frac{1}{2} \sum_{k} \tilde{\Psi}^\dagger_{k,Y+1} \left( -\frac{t_J}{2} \sigma_z \right) \tilde{\Psi}_{k,Y} + h.c., \nonumber \\
    &\approx i\frac{\delta_J}{4} \sum_{k} (\gamma^-_{k,Y+1})^\dagger \gamma^+_{k,Y} + h.c. \;,\label{appEq:junctionCoupling}
\end{align}
with tunneling alone, where we have projected $H_J$ onto the subspace of low-energy junction modes, in the second line. From the viewpoint of low-energy modes, this coupling only differs from $H^\text{direct}_J$ by a factor $1/2$. The resulting low-energy single-particle spectrum (spectrum of $H_{JL} + H_{JR} + H_J$) is illustrated in Fig.~\ref{fig:josephsonEffect}. It exhibits a clear $4\pi$ periodicity in $\delta\chi$, as expected. The key features of this spectrum are the energy crossings appearing at odd values of $\delta\chi/\pi$.

We emphasize that the details of the junction coupling are mostly irrelevant for our purposes: Two energy crossings generally appear when varying $\delta\chi$ by $4\pi$, for arbitrary couplings $H'_J \neq H_J$. The strict $4\pi$ periodicity of the Josephson current requires $H'_J$ to have a constant overlap with $H_J$ when projected onto the subspace of low-energy junction modes. As mentioned in the main text, $4\pi$ periodic Josephson effects have been identified in various setups based on integer topological superconductors~\cite{Kitaev2001,Fu2008,Fu2009}. Our results show that they can be understood as manifestations of the same bulk pump $\delta\chi$.

\section{Generalized coupled-wire description of integer and fractional ($l \geq 1$) topological phases}
\label{sec:generalizedCoupledWireConstruction}

In previous Appendices, we have relied on noninteracting tight-binding models to demonstrate the pumping effects of bulk flux quanta explicitly. We have focused on two types of integer (short-range entangled) 2D topological phases amenable to a convenient coupled-wire description: quantum Hall and topological superconducting phases (belonging to classes A and D of standard classifications~\cite{Zirnbauer1996,Altland1997}, respectively). In the following, we generalize this coupled-wire approach to describe interacting analogs of these phases, focusing on fractional (long-range entangled) variants thereof. We follow the formalism of Refs.~\cite{Neupert2014,Huang2016}, based on Refs.~\cite{Kane2002,Teo2014} and on a formulation of Abelian bosonization by Haldane~\cite{Haldane1995}.

\subsection{Main framework}
\label{sec:mainFramework}

As in previous Appendices, we consider a cylindrical system consisting of $N_y$ coupled identical wires (rings) oriented along the $x$ direction, with periodic boundary conditions and indices $y = 1, \ldots, N_y$ corresponding to wire positions in the $y$ direction. We assume that individual wires have $N_\nu$ internal fermionic degrees of freedom --- indexed by $\nu = 1, \ldots, N_\nu$ --- forming a total of $N_y N_\nu$ fermionic fields described by creation and annihilation operators $\psi_j^\dagger(x)$ and $\psi_j(x)$, respectively [where $j = 1, \ldots, N_y N_\nu$ corresponds to the composite index $(y, \nu)$]. Fields are collected into a vector $\boldsymbol{\Psi}(x) = [\psi_1(x),\ldots,\psi_{N_y N_\nu}(x)]^T$, and we omit their explicit time dependence. We assume that inter-wire couplings are weak as compared to couplings within individual wires, so that the latter can be regarded as Luttinger liquids with $N_\nu$ fermionic channels, contributing to a total of $N_y N_\nu$ channels. This set of channels can be bosonized following the conventional prescriptions of Abelian bosonization. 
Specifically, one can define a vector $\boldsymbol{\Phi}(x)$ of Hermitian fields $\phi_j(x)$ related to the original fermionic fields by the Matthis-Mandelstam formula
\begin{equation} \label{appEq:psiPhiRelation}
    \boldsymbol{\Psi}(x) = \,: \exp \left[ i K \boldsymbol{\Phi}(x) \right] :\,,
\end{equation}
where ``$:\ldots:$'' denotes normal ordering. Here, $K$ is a symmetric integer-valued $N_y N_\nu \times N_y N_\nu$ matrix which is block-diagonal [$K_{jk} \equiv K_{(y,\nu)(y',\nu')} = \delta_{yy'} \mathcal{K}_{\nu,\nu'}]$, since wires are identical. The fields $\phi_j(x)$ satisfy the equal-time bosonic Kac-Moody algebra~\cite{Haldane1995}
\begin{align}
    \left[ \phi_j(x), \phi_k(x') \right] = -i \pi & \left[ K_{jk}^{-1} \sgn(x-x') \right. \nonumber \\
    & \left. \phantom{[} + K_{jj'}^{-1} L_{j'k'} K_{k'k}^{-1} \right], \label{appEq:phiCommutationRelations}
\end{align}
and periodic boundary conditions
\begin{equation} \label{appEq:PhiBCs}
    K \boldsymbol{\Phi}(x+N_x) = K \boldsymbol{\Phi}(x) + 2\pi \mathbf{n},
\end{equation}
where $\mathbf{n}$ is a vector of integers, and $N_x$ is the length of the system in the $x$ direction. The matrix $L$ is antisymmetric, defined as $L_{jk} = \sgn(j-k)(K_{jk} + q_j q_k)$, where $q_j$ is the charge of fermions in channel $j$, and $\sgn(j-k) = 0$ when $j = k$. The term $K_{jj'}^{-1} L_{j'k'} K_{k'k}^{-1}$ in Eq.~\eqref{appEq:phiCommutationRelations}, sometimes called ``Klein factor''~\cite{Haldane1995}, ensures that vertex operators such as the $\psi_j(x)$ in Eq.~\eqref{appEq:psiPhiRelation} obey proper mutual commutation relations. We consider fermions with unit charge $q_j = 1$ for all $j$ (in natural units $\hbar = e = c = 1$).

We first examine the situation without flux variations $\delta\Phi$ or $\delta\chi$ [corresponding to $\delta A(y) = 0$ in the main text and in previous Appendices]. In that case, the many-body Hamiltonian describing the low-energy effective field theory of the coupled-wire array can be expressed in the generic form
\begin{align}
    H & = H_0 + \sum_{c \in \mathcal{C}} H_c, \label{appEq:H} \\
    H_0 & = \int dx [\partial_x \boldsymbol{\Phi}(x)]^T V(x) [\partial_x \boldsymbol{\Phi}(x)], \label{appEq:H0} \\
    H_c & = \int dx \alpha_c(x) e^{-i\beta_c(x)} \prod_j \psi_j^{v_{c,j}}(x) + h.c. \nonumber \\ 
    & = 2 \int dx \alpha_c(x) :\cos\left[ \mathbf{v}_c^T K \boldsymbol{\Phi}(x) + \beta_c(x) \right]:. \label{appEq:Hc}
\end{align}
The Hamiltonian term $H_0$, which is quadratic in the fields, describes two types of contributions in individual wires: one-body terms, and two-body density-density interactions (or ``forward-scattering'' terms~\cite{Kane2002,Teo2014}). The corresponding $N_y N_\nu \times N_y N_\nu$ matrix $V(x)$ is real symmetric and block-diagonal: $V_{jk} \equiv \delta_{yy'} \mathcal{V}_{\nu,\nu'}$ (no density-density interactions between wires, for simplicity). The Hamiltonian terms $H_c$, which are typically not quadratic in the fields, describe all other types of couplings between fermionic channels. We denote the relevant set of couplings by $\mathcal{C}$, and represent each coupling $c \in \mathcal{C}$ by a vector $\mathbf{v}_c$ with elements $v_{c,j} \in \{ -1, 0, 1 \}$, with the convention that $\psi_j^{v_{c,j}}(x) \equiv \psi_j^\dagger(x)$ for $v_{c,j} = -1$. The coupling amplitude and phase are defined by real quantities $\alpha_c(x) > 0$ and $\beta_c(x)$, respectively.

We remark that a macroscopic background transverse flux $\chi_0$ may be required to enable the couplings $H_c$. This is the case in the quantum Hall phases discussed in the main text and in Appendix~\ref{sec:tightBindingQH}, in particular, where $\chi_0$ controls the filling factor, or the relative momentum shift between fermionic degrees of freedom in distinct wires. In the integer case, $\chi_0$ ensures that right movers in wire $y$ are resonant with left movers in wire $y+1$, enabling their direct coupling (see Fig.~\ref{fig:coupledWireIQH}b). In topological superconducting phases, no background flux is required. In the integer case, due to particle-hole symmetry, the desired left and right movers can be coupled directly by a combination of tunneling and superconducting pairing, as discussed in the main text and in Appendix~\ref{sec:tightBindingSC} (Fig.~\ref{fig:coupledWireSC}b). In quantum Hall phases, couplings preserve the total fermion number (such that $\sum_j v_{c,j} = 0$), whereas couplings preserve the total fermion parity (such that $\sum_j v_{c,j} = 0$ modulo $2$) in topological superconductors.

\subsubsection*{Basis transformations}

Before constructing the phases of interest, we discuss several generic properties of the above theory. We first note that it is invariant under basis transformations of the form
\begin{align}
    \tilde{\boldsymbol{\Phi}}(x) & = G^{-1} \boldsymbol{\Phi}(x), \label{appEq:basisTransformation1} \\
    \tilde{K} & = G^T K G, \\
    \tilde{L} & = G^T L G, \\
    \tilde{q} & = G^T q, \label{appEq:basisTransformationq} \\
    \tilde{V}(x) & = G^T V(x) G, \\
    \tilde{\mathbf{v}}_c & = G^{-1} \mathbf{v}_c, \label{appEq:basisTransformation2}
\end{align}
where $G$ is an invertible integer-valued $N_y N_\nu \times N_y N_\nu$ matrix, and $q$ is a vector of charges $q_j$. The Hamiltonian defined in Eqs.~\eqref{appEq:H}-\eqref{appEq:Hc} remains of the same form under the transformation $G$, with $V(x), \mathbf{v}_c \to \tilde{V}(x), \tilde{\mathbf{v}}_c$, and transformed fields $\tilde{\boldsymbol{\Phi}}(x)$ obeying the same algebra as in Eq.~\eqref{appEq:phiCommutationRelations}, with $K, L \to \tilde{K}, \tilde{L}$. The corresponding vertex operators $\tilde{\boldsymbol{\Psi}}(x) = \,: \exp[i \tilde{K} \tilde{\boldsymbol{\Phi}}(x)] :$ are distinct from the original fermionic fields defined in Eq.~\eqref{appEq:psiPhiRelation}, namely,
\begin{equation} \label{}
    \tilde{\boldsymbol{\Psi}}(x) = \,: \exp \left[ i G^T K \boldsymbol{\Phi}(x) \right] :\, \neq \boldsymbol{\Psi}(x).
\end{equation}
We remark that the matrix $K$ is sometimes absorbed in the definition of the fields $\boldsymbol{\Phi}(x)$ in Eq.~\eqref{appEq:psiPhiRelation} [i.e., $K \boldsymbol{\Phi}(x) \to \boldsymbol{\Phi}(x)$], as in Refs.~\cite{Kane2002,Teo2014}. This corresponds to a basis transformation $G = K^{-1}$.

\subsubsection*{Requirements for a bulk spectral gap}

The Hamiltonian defined in Eqs.~\eqref{appEq:H}-\eqref{appEq:Hc} describes the competition between two types of terms: the Hamiltonian $H_0$ of uncoupled wires, which supports $N_y N_\nu$ gapless modes, and the inter-wire couplings $H_c$, which gap some (if not all) of these modes. Starting from the fixed point corresponding to $H_0$ alone, one can introduce couplings $H_c$ within a specific symmetry class, and use renormalization-group theory to identify coupling vectors $\mathbf{v}_c$ that make the system flow to a fixed point corresponding to a gapped phase with robust (topological) gapless edge modes. Following Ref.~\cite{Neupert2014}, we do not solve such a renormalization problem but focus, instead, on the strong-coupling limit defined by $\alpha_c(x) \to \infty$ in Eq.~\eqref{appEq:Hc}. We assume that this limit corresponds to a stable point which can be reached from $H_c = 0$ without getting trapped in intermediary fixed points along the renormalization-group flow.

Although $H_0$ is negligible in the strong-coupling limit, quantum fluctuations due to commutation relations between fields [Eq.~\eqref{appEq:phiCommutationRelations}] do not necessarily allow us to find a solution which minimizes the energy of all couplings $H_c$ separately. For each $H_c$, energy minimization requires the phase gradient $\partial_x \boldsymbol{\Phi}(x)$ to be locked to $\partial_x \beta_c(x)$ at all times, i.e., we must have
\begin{equation} \label{appEq:lockingCondition}
    \partial_x \left[ \mathbf{v}_c^T K \boldsymbol{\Phi}(t, x) + \beta_c(x) \right] = f_c(t, x),
\end{equation}
for some real function $f_c(t, x)$~\cite{Haldane1995} (where we have briefly restored the explicit time dependence of fields). For this locking to survive over time, $\mathbf{v}_c^T K \partial_x \boldsymbol{\Phi}(t, x)$ must be a constant of motion, i.e., $[\mathbf{v}_c^T K \partial_x \boldsymbol{\Phi}(x), H_{c'}]$ must vanish for all couplings $c' \in \mathcal{C}$ including $c' = c$. As detailed in Ref.~\cite{Neupert2014}, this leads to the so-called ``Haldane criterion''
\begin{equation} \label{appEq:HaldaneCriterion}
    \mathbf{v}_c^T K \mathbf{v}_{c'} = 0.
\end{equation}
When Eq.~\eqref{appEq:HaldaneCriterion} is satisfied for all $c, c' \in \mathcal{C}$, couplings $H_c$ are compatible with each other, i.e., they all satisfy a locking condition as in Eq.~\eqref{appEq:lockingCondition}. In that case, inter-wire couplings $H_c$ each gap out a \emph{distinct} pair of gapless modes, removing the latter from the low-energy theory of the system. We will be interested in topological phases where the only remaining gapless modes are edge modes, which cannot be gapped by quasilocal perturbations (within a relevant symmetry class) because of the spatial separation between edges.

\subsubsection*{Models of interest}

The above coupled-wire formalism has been used to construct a variety of gapped topological phases with robust low-energy gapless edge modes (see, e.g., Refs.~\cite{Kane2002,Teo2014,Neupert2014,Huang2016}). Models are generically specified by: (i) the number and type of degrees of freedom in each wire, (ii) the symmetries of the system, and (iii) the inter-wire couplings. In the following, we detail explicit models for the two types of phases used in the main text (and in Appendices~\ref{sec:tightBindingQH} and~\ref{sec:tightBindingSC}) to illustrate bulk pumping effects: integer and fractional quantum Hall phases, which do not require any specific symmetry (symmetry class A of conventional classifications~\cite{Zirnbauer1996,Altland1997}), and integer and fractional topological superconductors, which are protected by particle-hole (PH) symmetry (symmetry class D). In each case, we identify a set of inter-wire couplings that (i) belong to the desired symmetry class, (ii) act quasilocally, corresponding to short-range scatterings or interactions, and (iii) is maximal, in the sense that the corresponding coupling vectors $\mathbf{v}_c$ form a (typically non-unique) maximal set of linearly independent vectors satisfying the Haldane criterion [Eq.~\eqref{appEq:HaldaneCriterion}]. Our goal is to identify a set of couplings that gaps all modes in the bulk while leaving gapless edge modes which cannot be gapped, as the set is maximal.

\subsection{Explicit model for integer and fractional quantum Hall insulators}
\label{sec:bosonizedPictureQH}

We first examine the case of quantum Hall phases, which do not rely on any symmetry besides $U(1)$ charge conservation. As in Appendix~\ref{sec:tightBindingQH}, we start from an array of $N_y$ uncoupled identical wires supporting each $N_\nu = 2$ internal degrees of freedom, namely: left- and right-moving spinless fermionic modes at the Fermi level (recall the inset of Fig.~\ref{fig:coupledWireIQH}a). In the above framework, we describe these $N_y N_\nu$ fermionic fields or channels $\boldsymbol{\Psi}(x)$ by chiral bosonic fields $\boldsymbol{\Phi}(x)$ defined by Eq.~\eqref{appEq:psiPhiRelation}, with
\begin{equation} \label{}
    K = \mathbb{I}_{N_y} \otimes \diag(-1, +1),
\end{equation}
where $-1$ ($+1$) corresponds to left (right) movers, and $\mathbb{I}_{N_y}$ is the $N_y \times N_y$ identity matrix. The fields $\boldsymbol{\Phi}(x)$ obey commutation relations given by Eq.~\eqref{appEq:phiCommutationRelations} (with unit charge $q_j = 1$ for all fermionic channels). In the absence of density-density interactions between fermionic channels, uncoupled wires are described by the Hamiltonian $H_0$ in Eq.~\eqref{appEq:H0}, with diagonal matrix
\begin{equation} \label{appEq:VQH}
    V(x) \equiv V = \mathbb{I}_{N_y} \otimes v_F \diag(-1, 1),
\end{equation}
where $v_F$ is the Fermi velocity of noninteracting wires. The only effect of density-density interactions between channels is to renormalize $v_F \to \widetilde{v_F}$.

To couple wires and generate a gapped phase, we introduce couplings which, as discussed above, satisfy three requirements: (i) They preserve the symmetries [here, $U(1)$ charge conservation], (ii) they are quasilocal, and (iii) they form a maximal set satisfying the Haldane criterion [Eq.~\eqref{appEq:HaldaneCriterion}]. Focusing on couplings acting on nearest-neighboring wires, for simplicity, the only possible choice of coupling vectors is, up to a global integer factor,
\begin{equation} \label{appEq:couplingsA}
    \mathbf{v}_c \equiv \mathbf{v}_{y,y+1} = (0, 0 | \ldots | -l_-, l_+ | -l_+, l_- | \ldots | 0, 0)^T, 
\end{equation}
where $l_\pm \equiv (l \pm 1)/2$, and $l$ is the odd positive integer used in the main text (vertical lines separate elements from distinct wires). The corresponding inter-wire coupling Hamiltonian $H_c \equiv H_{y,y+1}$ is given by Eq.~\eqref{appEq:Hc}, which parallels Eq.~\eqref{appEq:interwireCouplingQH} of Appendix~\ref{sec:tightBindingQH} for the integer case $l = 1$ [setting $\alpha_c(x) = -t_\perp/2$ and $\beta_c(x) = 0$]. The $N_y-1$ couplings $H_{y,y+1}$ gap out $2(N_y-1)$ of the $2N_y$ degrees of freedom of the system. The two remaining gapless modes are located at the edges, and are topologically protected. Indeed, the only additional coupling which could satisfy the Haldane criterion is
\begin{equation} \label{}
    \mathbf{v}_0 = ( -l_+, l_- | 0, 0 | \ldots | 0, 0 | -l_-, l_+ )^T,
\end{equation}
which is highly nonlocal, with support at both edges. To understand the properties of gapless edge modes, we perform a basis transformation $\tilde{\boldsymbol{\Phi}}(x) = G^{-1} \boldsymbol{\Phi}(x)$ as in Eqs.~\eqref{appEq:basisTransformation1}-\eqref{appEq:basisTransformation2}, with
\begin{align} \label{}
    G^{-1} & = \mathbb{I}_{N_y} \otimes \frac{1}{l} \left( \begin{array}{cc}
        l_+ & l_- \\
        l_- & l_+
    \end{array} \right), \\
    G & = \mathbb{I}_{N_y} \otimes \left( \begin{array}{cc}
        l_+ & -l_- \\
        -l_- & l_+
    \end{array} \right).
\end{align}
The relevant inter-wire coupling vectors become
\begin{align}
    \tilde{\mathbf{v}}_{y,y+1} & = G^{-1} \mathbf{v}_{y,y+1} = (0, 0 | \ldots | 0, 1 | -1, 0 | \ldots | 0, 0)^T, \nonumber \\
    \tilde{\mathbf{v}}_0 & = G^{-1} \mathbf{v}_0 = (-1, 0 | 0, 0 | \ldots | 0, 0 | 0, 1)^T, \label{appEq:couplingVectorsTildeA}
\end{align}
with transformed matrices $K, L, V$ of the form
\begin{align}
    \tilde{K} & = G^T K G = \mathbb{I}_{N_y} \otimes \diag(-l, l), \label{appEq:KMatrixTildeQH} \\
    \tilde{L} & = G^T L G = \mathbb{I}_{N_y} \otimes (-i \sigma_y)l + \Sigma_{2N_y}, \label{appEq:LMatrixTildeQH} \\
    \tilde{V} & = G^T V G = \mathbb{I}_{N_y} \otimes v_F \diag(-l, l), \label{appEq:VMatrixTildeQH}
\end{align}
where $\sigma_y$ is the standard Pauli matrix, and $\Sigma_n$ is an antisymmetric $n \times n$ matrix with elements $(\Sigma_n)_{jk} \equiv (\Sigma_n)_{(y,\nu)(y',\nu')} = \sgn(y-y')$. The modified fields $\tilde{\boldsymbol{\Phi}}(x) = G^{-1} \boldsymbol{\Phi}(x)$ satisfy commutation relations given by Eq.~\eqref{appEq:phiCommutationRelations}, with modified Klein factors $\tilde{L} = G^T L G$ giving a factor $l$ in the commutator $[\tilde{\phi}_j(x), \tilde{\phi}_k(x')]$, when fields $\tilde{\phi}_j(x)$ and $\tilde{\phi}_k(x')$ belong to the same wire.

Equations~\eqref{appEq:couplingVectorsTildeA}-\eqref{appEq:VMatrixTildeQH} determine the low-energy theory of the system: Eq.~\eqref{appEq:couplingVectorsTildeA} shows that the remaining pair of gapless modes corresponds to the edge fields $\tilde{\phi}_1(x) \equiv \varphi_1(x) \equiv \varphi_L(x)$ and $\tilde{\phi}_{2N_y}(x) \equiv \varphi_2(x) \equiv \varphi_R(x)$, which are not affected by inter-wire couplings. The algebra of these fields is determined by Eqs.~\eqref{appEq:KMatrixTildeQH} and~\eqref{appEq:LMatrixTildeQH}, i.e., transformed fields $\tilde{\phi}_j(x)$ obey commutation relations in Eq.~\eqref{appEq:phiCommutationRelations} with $K, L \to \tilde{K}, \tilde{L}$, such that
\begin{align} \label{appEq:KacMoodyLevell}
    \left[ \varphi_j(x), \varphi_k(x') \right] & = -i \pi \left[ (K_\text{eff}^{-1})_{jk} \sgn(x-x') \right. \nonumber \\
    & \phantom{=} \left. + (K_\text{eff}^{-1})_{jj'} (L_\text{eff})_{j'k'} (K_\text{eff}^{-1})_{k'k} \right],
\end{align}
where we have defined
\begin{align}
    K_\text{eff} & = \diag(-l, l), \label{appEq:KEffectiveA} \\
    L_\text{eff} & = -i \sigma_y.
\end{align}
The corresponding low-energy Hamiltonian is described by $H_0$ in Eq.~\eqref{appEq:H0}, with $V(x) \to \tilde{V}$ given by Eq.~\eqref{appEq:VMatrixTildeQH}:
\begin{equation} \label{appEq:edgeHamiltonianApp}
    H_\sigma = \frac{v_Fl}{4\pi} \int dx (\partial_x \varphi_\sigma)^2,
\end{equation}
where $\sigma = -/+$ indexes the left and right edges of the system, respectively [with $\varphi_-(x) \equiv \varphi_1(x)$ and $\varphi_+(x) \equiv \varphi_2(x)$]. We thus recover the low-energy theory 
given by Eqs.~\eqref{eq:edgeHamiltonianQH}-\eqref{eq:periodicBCPhi} in the main text, with $v_F \to v_\sigma$ in the presence of density-density interactions between fermionic channels [see Eq.~\eqref{appEq:VQH}]. 

States in the above coupled-wire model are topologically equivalent to Laughlin states with index $l$ in the Abelian hierarchy of fractional quantum Hall phases~\cite{Laughlin1983,Halperin1983,Jain1989}. The commutation relations in Eq.~\eqref{appEq:KacMoodyLevell} (Kac-Moody algebra at level $l$) imply that the low-energy theory supports quasiparticle edge excitations (Laughlin quasiparticles) with fractional charge $e/l$ and fractional phase $\pi/l$ under spatial exchange~\cite{Wen1990,Wen1991,Stone1991,Frohlich1991}. Indeed, quasiparticle edge excitations are created by vertex operators
\begin{equation} \label{appEq:vertexOperatorsA}
    (\Psi_j^\text{qp})^\dagger(x) = \,: \exp[-i \varphi_j(x)] :\,.
\end{equation}
The corresponding charge can be identified by examining the commutator $[Q_j, (\Psi_j^\text{qp})^\dagger(x)]$, where $Q_j$ is the total charge along the edge $j$. Here we have $Q_j = \tilde{q}_j N_j$, where $\tilde{q}_j = -\sigma$ 
is the charge associated with $\varphi_j(x)$ [Eq.~\eqref{appEq:basisTransformationq}], and $N_j$ is the density integrated along the edge:
\begin{equation} \label{}
    N_j = \frac{1}{2\pi} \int_0^{L_x} dx \partial_x \varphi_j(x),
\end{equation}
which is a conserved quantity, as $\partial_t N_j = 0$ due to boundary conditions [Eq.~\eqref{appEq:PhiBCs}]. Since $[N_j, \varphi_k(x)] = -i (K_\text{eff}^{-1})_{jk}$ [from Eq.~\eqref{appEq:KacMoodyLevell} with $\partial_x \sgn(x-x') = 2\delta(x-x')$], we find
\begin{align} \label{}
    \left[ Q_j, (\Psi_j^\text{qp})^\dagger(x) \right] &= -\tilde{q}_j (K_\text{eff}^{-1})_{jk} (\Psi_k^\text{qp})^\dagger(x) \nonumber \\
    &= \frac{e}{l} (\Psi_j^\text{qp})^\dagger(x),
\end{align}
with $K_\text{eff}$ given by Eq.~\eqref{appEq:KEffectiveA}. This verifies that $(\Psi_j^\text{qp})^\dagger(x)$ creates quasiparticles with charge $e/l$. The exchange statistics of these quasiparticles can be derived from Eq.~\eqref{appEq:KacMoodyLevell} using the Baker-Campbell-Hausdorff formula:
\begin{align} \label{}
\begin{split}
    & \Psi_j^\text{qp}(x) \Psi_k^\text{qp}(x') = \Psi_k^\text{qp}(x') \Psi_j^\text{qp}(x) \ldots \\
    & \quad \times \exp \left[ -i \pi (K_\text{eff}^{-1})_{jk} \sgn(x-x') \right. \\
    & \phantom{\quad \times \exp[} \left. - \pi (K_\text{eff}^{-1})_{jj'} (\sigma_y)_{j'k'} (K_\text{eff}^{-1})_{k'k} \right].
\end{split}
\end{align}
This confirms that quasiparticles at one edge ($j = k$) exhibit a phase $\pi/l$ under spatial exchange (corresponding to Abelian anyons, in the fractional case $l > 1$).



When introducing flux variations $\delta\Phi, \delta\chi$ as described in the main text, corresponding to gauge-field variations $\delta A_j$ at each of the edges $j$ [see Eq.~\eqref{eq:gaugeChoice1}], the above low-energy edge theory changes according to the standard prescriptions of minimal coupling, i.e.,
\begin{equation} \label{appEq:minimalCouplingChiralBosonicFields}
    \varphi_j(x) \to \varphi_j(x) - \frac{e}{l} \int_0^x dx' \delta A_j(x'),
\end{equation}
as in Eq.~\eqref{eq:minimalCouplingChiralBosonicFields} of the main text. This can be understood by remembering that fermions with unit charge are described by operators such as $\Psi_j^\text{f} = :\exp \left[ i (K_\text{eff})_{jk} \varphi_k(x) \right]:$, with $K_\text{eff} = \diag(-l, l)$ [Eq.~\eqref{appEq:KEffectiveA}].

\subsection{Explicit model for integer and fractional topological superconductors}
\label{sec:bosonizedPictureSC}

To construct the topological superconducting phases with Majorana gapless edge theory considered in the main text, we start from the above coupled-wire array in symmetry class A (quantum Hall insulators), and add superconductor-induced pairings, i.e., couplings that (i) conserve the fermion-number parity instead of the total fermion number, and (ii) preserve particle-hole symmetry (PHS), thereby promoting the system to symmetry class D~\cite{Zirnbauer1996,Altland1997}. As in the previous section, our construction follows along the lines of Ref.~\cite{Neupert2014}.

To be able to describe superconducting pairings, we first move to a Bogoliubov de-Gennes (BdG) picture where particles and holes in each wire are regarded as independent, which artificially doubles the number of internal degrees of freedom. Explicitly, we consider the fermionic fields $\psi_j(x)$ and $\psi_j^\dagger(x)$ as independent, and collect them into a doubled vector (Nambu spinor) $\boldsymbol{\Psi}(x)$. The vector $\boldsymbol{\Phi}(x)$ of bosonic fields is similarly extended (doubled), ensuring that Eq.~\eqref{appEq:psiPhiRelation} still holds. Particles and holes are not truly independent, however, and the relation between $\psi_j(x)$ and $\psi_j^\dagger(x)$ implies the existence of an ``emergent'' PHS for physical operators in the BdG or Nambu representation. Explicitly, the subspace of physical operators is identified by the ``reality condition''
\begin{equation} \label{appEq:realityConditionPsi}
    \Pi \boldsymbol{\Psi} \Pi^\dagger = \boldsymbol{\Psi}^\dagger,
\end{equation}
or, equivalently,
\begin{equation} \label{appEq:realityConditionPhi}
    \Pi \boldsymbol{\Phi} \Pi^\dagger = -\boldsymbol{\Phi}.
\end{equation}
where $\Pi$ is the unitary many-body operator representing the relevant PHS.
The action of PHS on the fields can be represented in the generic form
\begin{equation} \label{appEq:PHSPsi}
    \Pi \boldsymbol{\Psi} \Pi^\dagger = P_\Pi \boldsymbol{\Psi} e^{i \pi D_\Pi},
\end{equation}
or, equivalently,
\begin{equation} \label{appEq:PHSPhi}
    \Pi \boldsymbol{\Phi} \Pi^\dagger = P_\Pi \boldsymbol{\Phi} + \pi K^{-1} \mathbf{d}_\Pi,
\end{equation}
where $P_\Pi$ is a $N_y N_\nu \times N_y N_\nu$ permutation matrix (such that $P_\Pi^{-1} = P_\Pi^T$) describing the exchange of particles and holes in individual wires, and $\mathbf{d}_\Pi$ is an integer-valued vector describing the corresponding phase (if any), with $D_\Pi \equiv \diag(\mathbf{d}_\Pi)$. 
The system is particle-hole symmetric whenever its Hamiltonian satisfies
\begin{equation} \label{}
    \Pi H \Pi^\dagger = H.
\end{equation}
Remembering the form of $H$ [Eq.~\eqref{appEq:H}] and using Eq.~\eqref{appEq:PHSPhi}, we obtain the following conditions for PHS:
\begin{align}
    P_\Pi^{-1} V(x) P_\Pi & = V(x). \label{appEq:PHSV} \\
    P_\Pi^{-1} K P_\Pi & = K, \label{appEq:PHSK} \\
    P_\Pi \mathbf{v}_c & = \pm \mathbf{v}_c, \label{appEq:PHSvc} \\
    \beta_c(x) & = \pm \left[ \beta_c(x) + \pi \mathbf{v}_c^T P_\Pi \mathbf{d}_\Pi \right] (\text{mod } 2\pi), \label{appEq:PHSbeta}
\end{align}
with the same choice of sign in the last two lines.

We now construct an explicit model for integer and fractional topological superconductors, which, in the integer case, reduces to the tight-binding model presented in Appendix~\ref{sec:tightBindingSC}. We start from an array of $N_y$ uncoupled wires supporting each a pair of left- and right-moving spinless fermionic modes which, in Nambu (doubled) space, translates as $N_\nu = 4$ degrees of freedom. The relevant matrix $K$ reads
\begin{equation} \label{}
    K = \mathbb{I}_{N_y} \otimes \diag(-1, +1, +1, -1),
\end{equation}
where $-1, +1, +1, -1$ respectively correspond to left- and right-moving particles, and right- and left-moving holes. In this picture, PHS is represented by
\begin{equation} \label{}
    P_\Pi = \mathbb{I}_M \otimes \left( \begin{array}{cccc}
        0 & 0 & 0 & 1 \\
        0 & 0 & 1 & 0 \\
        0 & 1 & 0 & 0 \\
        1 & 0 & 0 & 0
    \end{array} \right), \quad \mathbf{d}_\Pi = \mathbf{0},
\end{equation}
where $\mathbf{d}_\Pi = \mathbf{0}$ reflects the spinless nature of particles (and holes). Using Eq.~\eqref{appEq:PHSPhi}, the reality condition that must be imposed in Nambu space [Eq.~\eqref{appEq:realityConditionPhi}] takes the form $P_\Pi \boldsymbol{\Phi} = -\boldsymbol{\Phi}$, such that
\begin{equation} \label{appEq:appliedRealityConditionPhi}
    \boldsymbol{\Phi}^T = (\ldots \phi_{y,1}, \phi_{y,2}, \phi_{y,3} = -\phi_{y,2}, \phi_{y,4} = -\phi_{y,1}),
\end{equation}
where we have omitted explicit position and time dependences. The fields $\phi_{y,3}$ and $\phi_{y,4}$, which are regarded as independent degrees of freedom in Nambu space, are thus directly related to $\phi_{y,1}$ and $\phi_{y,2}$. More importantly, the reality condition $P_\Pi \boldsymbol{\Phi} = -\boldsymbol{\Phi}$ implies that inter-wire couplings must satisfy $P_\Pi \mathbf{v}_c = -\mathbf{v}_c$ in Eq.~\eqref{appEq:PHSvc} to be physical. Indeed, couplings depend on fields via $\mathbf{v}_c^T K \boldsymbol{\Phi}(x)$  [Eq.~\eqref{appEq:Hc}], which vanishes when $P_\Pi \mathbf{v}_c = +\mathbf{v}_c$ and $P_\Pi \boldsymbol{\Phi} = -\boldsymbol{\Phi}$. Without loss of generality, we can thus introduce a set of fictitious local couplings $\mathbf{v}_c \equiv \mathbf{v}_y^f$ which satisfy $P_\Pi \mathbf{v}_c = +\mathbf{v}_c$ and, hence, ``gap out'' unphysical degrees of freedom:
\begin{equation} \label{appEq:couplingVectorsFictitious}
    \mathbf{v}_y^f = (0, 0, 0, 0 | \ldots | -1, 1, 1, -1 | \ldots | 0, 0, 0, 0)^T,
\end{equation}
up to an integer factor. Note that $(\mathbf{v}_y^f)^T K \mathbf{v}_y^f = 0$, as required by the Haldane criterion [Eq.~\eqref{appEq:HaldaneCriterion}].

As in Appendix~\ref{sec:tightBindingSC}, we start from an array of uncoupled gapless wires supporting a pair of PH symmetric chiral modes. In analogy with Eq.~\eqref{appEq:linearizedWireHamMomSpaceSC}, the Hamiltonian $H_0$ of uncoupled wires takes the form of Eq.~\eqref{appEq:H0}, with
\begin{equation} \label{appEq:VSC}
    V(x) \equiv V = \mathbb{I}_{N_y} \otimes \frac{\Delta}{2} \diag(-1, 1, 1, -1),
\end{equation}
where $\Delta$ is the velocity of chiral modes in individual wires. The factor $1/2$ in $\mathbf{v}_{y,y+1}$ compensates for the doubling of degrees of freedom in Nambu space. To gap the system in a way that generates the topological superconducting phase of interest, with a pair of chiral gapless modes at the edges, we introduce inter-wire couplings
\begin{widetext}
\begin{equation} \label{appEq:couplingVectorsSC}
    \mathbf{v}_{y,y+1} = \frac{1}{2} (0, 0, 0, 0 | \ldots | -l_-, l_+, -l_+, l_- | -l_+, l_-, -l_-, l_+ | \ldots | 0, 0, 0, 0)^T,
\end{equation}
\end{widetext}
acting on nearest-neighboring wires, for simplicity. Here, $l_\pm \equiv (l \pm 1)/2$ with odd integer $l > 0$, as in the quantum Hall case. In the ``integer'' case $l = 1$ (where $l_+ = 1$ and $l_- = 0$), $\mathbf{v}_{y,y+1}$ corresponds to the coupling $H_{y,y+1}$ introduced in Eq.~\eqref{appEq:interwireCouplingSC} of Appendix~\ref{sec:tightBindingSC}: It describes a direct coupling between the PH symmetric right-moving mode of wire $y$, and the PH symmetric left-moving mode of wire $y+1$. Since $P_\Pi \mathbf{v}_{y,y+1} = -\mathbf{v}_{y,y+1}$, the corresponding coupling phase must be real, i.e., $\beta_c(x) \equiv \beta_{y,y+1} = 0$ or $\pi$ [see Eq.~\eqref{appEq:PHSbeta}].

Together, the inter-wire couplings $\mathbf{v}_y^f$ and $\mathbf{v}_{y,y+1}$ gap out $2N_y + 2(N_y-1) = 4N_y-2$ of the $4N_y$ chiral gapless modes of the system. The remaining two modes are topologically protected chiral edge modes. Indeed, the only coupling that could gap them while satisfying the Haldane criterion [Eq.~\eqref{appEq:HaldaneCriterion}] is, up to an integer factor,
\begin{align} \label{appEq:nonlocalEdgeCouplingD}
\begin{split}
    \mathbf{v}_0 = \frac{1}{2} ( &-l_+, l_-, -l_-, l_+ | 0, 0, 0, 0 | \ldots \\
    & \ldots | 0, 0, 0, 0 | -l_-, l_+, -l_+, l_-)^T.
\end{split}
\end{align}
To identify the nature of the remaining low-energy gapless edge theory, we perform a similar basis transformation $\tilde{\boldsymbol{\Phi}}(x) = G^{-1} \boldsymbol{\Phi}(x)$ as in the quantum Hall case, with
\begin{align} \label{}
    G^{-1} & = \mathbb{I}_M \otimes \frac{1}{l} \left( \begin{array}{cccc}
        l_+ & l_- & 0 & 0 \\
        l_- & l_+ & 0 & 0 \\
        0 & 0 & l_+ & l_- \\
        0 & 0 & l_- & l_+
    \end{array} \right), \\
    G & = \mathbb{I}_M \otimes \left( \begin{array}{cccc}
        l_+ & -l_- & 0 & 0 \\
        -l_- & l_+ & 0 & 0 \\
        0 & 0 & l_+ & -l_- \\
        0 & 0 & -l_- & l_+
    \end{array} \right).
\end{align}
The relevant matrices $K, L, V$ become
\begin{align}
    \tilde{K} & = G^T K G = \mathbb{I}_{N_y} \otimes \diag(-l, l, l, -l), \label{appEq:KMatrixTildeSC} \\
    \tilde{L} & = G^T L G = \mathbb{I}_{2N_y} \otimes (-i \sigma_y)l + \Sigma_{4N_y}, \label{appEq:LMatrixTildeSC} \\
    \tilde{V} & = G^T V G = \mathbb{I}_{N_y} \otimes \Delta \diag(-l, l, l, -l), \label{appEq:VMatrixTildeSC}
\end{align}
where we recall that $(\Sigma_n)_{jk} \equiv (\Sigma_n)_{(y,\nu)(y',\nu')} = \sgn(y-y')$. The nonlocal coupling $\mathbf{v}_0$ in Eq.~\eqref{appEq:nonlocalEdgeCouplingD} becomes
\begin{align} \label{appEq:couplingVectorsTildeD}
\begin{split}
    \tilde{\mathbf{v}}_0 = G^{-1} \mathbf{v}_0 = \frac{1}{2} ( &-1, 0, 0, 1 | 0, 0, 0, 0 | \ldots \\
    & \ldots | 0, 0, 0, 0 | 0, 1, -1, 0)^T.
\end{split}
\end{align}
Equations~\eqref{appEq:KMatrixTildeSC}-\eqref{appEq:couplingVectorsTildeD} show that the low-energy theory that remains after integrating out gapped bulk modes is described by the edge fields $\tilde{\phi}_1(x) \equiv \varphi_1(x)$, $\tilde{\phi}_4(x) \equiv \varphi_2(x)$, $\tilde{\phi}_{4N_y-2}(x) \equiv \varphi_3(x)$, and $\tilde{\phi}_{4N_y-1}(x) \equiv \varphi_4(x)$, with effective matrices $K, L, V$ of the form
\begin{align} 
    K_\text{eff} & = \diag( -l, -l, l, l), \label{appEq:KEffectiveD} \\
    L_\text{eff} & = \Sigma_4, \\ 
    V_\text{eff} & = \diag( -l, -l, l, l),
\end{align}
and PHS represented by
\begin{equation} \label{}
    P_{\Pi, \text{eff}} = \left( \begin{array}{cccc}
        0 & 1 & 0 & 0 \\
        1 & 0 & 0 & 0 \\
        0 & 0 & 0 & 1 \\
        0 & 0 & 1 & 0
    \end{array} \right), \quad \mathbf{d}_{\Pi, \text{eff}} = \mathbf{0}.
\end{equation}
The above low-energy theory resembles two PH symmetric ``copies'' of the low-energy gapless edge theory derived for quantum Hall phases [Eqs.~\eqref{appEq:couplingVectorsTildeA}-\eqref{appEq:VMatrixTildeQH}], supporting Laughlin quasiparticles with charge $e/l$. Here, however, physical degrees of freedom are artificially doubled, as we work in Nambu space. Therefore, the physical low-energy theory actually corresponds to ``half'' of these two copies, i.e., it describes quasiparticles that are \emph{equal superpositions of Laughlin quasiparticles and quasiholes}. The physical theory is recovered by imposing the reality condition defined in Eq.~\eqref{appEq:realityConditionPhi}, corresponding to the identification $\varphi_2(x) = -\varphi_1(x)$ and $\varphi_4(x) = - \varphi_3(x)$. When $l = 1$, low-energy quasiparticle excitations take the form of chiral Majorana fermions, corresponding to PH symmetric superpositions of chiral quasiparticles and quasiholes with \emph{unit} charge. This was shown explicitly in the tight-binding model presented in Appendix~\ref{sec:tightBindingSC}. When $l > 1$, instead, quasiparticle excitations are superpositions of Laughlin quasiparticles and quasiholes with \emph{fractional} charge $e/l$. A single-particle picture is not suitable in that case, as inter-wire couplings [Eq.~\eqref{appEq:couplingVectorsSC}] correspond to true interactions. We remark that similar superpositions of Laughlin quasiparticles and quasiholes have been used to construct bound states known as ``parafermions'', or fractionalized Majorana fermions (see, e.g., Refs.~\cite{Lindner2012,Cheng2012,Clarke2013}).


\bibliography{bibliography}

\end{document}